\documentclass[11pt]{article}

\usepackage{amsmath, hyperref}
\usepackage{amssymb}
\usepackage{mathrsfs}
\usepackage{graphicx}
\usepackage{cite}
\usepackage[left=2.1cm,right=2.1cm,top=1.2cm,bottom=1.3cm,includeheadfoot]{geometry}
\usepackage{indentfirst}
\usepackage{color}

\newcommand{\average}[1]{\langle {#1}\rangle}

\newcommand{\fluct}[1]{(\widehat{#1}-{#1})}

\newcommand{\moma}{\pi_{a}}
\newcommand{\momphi}{\pi_{\phi}}
\newcommand{\momv}{\pi_{\mathbf{k}}}
\newcommand{\momvsigma}{\pi_{\mathbf{k}}}
\newcommand{\aop}{\widehat{a}}
\newcommand{\phiop}{\widehat{\phi}}

\newcommand{\momaop}{\widehat{\pi}_{a}}
\newcommand{\momphiop}{\widehat{\pi}_{\phi}}

\newcommand{\vk}{v_{\mathbf{k}}}
\newcommand{\vksigma}{v_{\mathbf{k}}}

\newcommand{\vi}{v_{\mathbf{q}}}
\newcommand{\visigma}{v_{\mathbf{q}}}
\newcommand{\vj}{v_{\mathbf{r}}}
\newcommand{\vjsigma}{v_{\mathbf{r}}}

\newcommand{\momvi}{\pi_{\mathbf{q}}}
\newcommand{\momvisigma}{\pi_{\mathbf{q}}}
\newcommand{\momvj}{\pi_{\mathbf{r}}}
\newcommand{\momvjsigma}{\pi_{\mathbf{r}}}

\newcommand{\ham}{\mathcal{H}_{a}}
\newcommand{\deltaunodos}[2]{\Delta(#1 #2^2)}
\newcommand{\deltadosuno}[2]{\Delta(#1^2 #2)}
\newcommand{\deltauno}[1]{\Delta(#1)}
\newcommand{\deltados}[1]{\Delta( #1^2)}
\newcommand{\deltatres}[1]{\Delta( #1 ^3 )}

\begin{document}

\begin{center}
{\Large\bf Mode coupling
on a geometrodynamical quantization of an\\[5pt] inflationary universe}

\vskip 4mm

David Brizuela
\footnote{
E-mail address: david.brizuela@ehu.eus}
and
Irene de Le\'on
\footnote{ E-mail address: irene.deleonperla@gmail.com}

\vskip 5mm
{\sl Fisika Saila, Universidad del Pa\'is Vasco/Euskal Herriko Unibertsitatea (UPV/EHU),\\ Barrio Sarriena s/n, 48940 Leioa, Spain}

\end{center}

\begin{quotation}
\noindent
\textbf{Abstract.}
A geometrodynamical quantization of an inflationary universe is considered in order
to estimate quantum-gravity effects for the primordial perturbations.
Contrary to previous studies in the literature,
the back-reaction produced by all the modes of the system is included
in our computations.
Even if at a classical level the assumption
that every mode evolves independently provides a good estimate for the
dynamics, our results explicitly show that this is not the case when considering
quantum-gravity effects. More precisely, both the self-interaction,
as well as the back-reaction from other modes, provide a correction of
the same order of magnitude to the usual power spectrum as computed
in the approximation of quantum field
theory on classical backgrounds. In particular, these quantum-gravity
effects introduce certain characteristic scale-dependence on the
expression of the power spectrum.
\end{quotation}

\vskip 2mm

\noindent

\section{Introduction}

The unification between gravity and the quantum theory has been one of the great challenges of theoretical physics
during several decades. Apart from the many conceptual issues that this construction has to face,
the lack of experimental evidence makes this search even more difficult.
Indeed, quantum-gravity effects might only become measurable at energies of the order of the Planck scale,
which is completely out of reach for the current particle accelerators. Nonetheless,
it has been argued that the highly energetic inflationary phase of the primordial universe
might be an adequate place to look for such effects. 

In order to study the evolution of the fluctuations, as a first approximation, one usually considers
quantum field theory (QFT) on classical backgrounds. In this context perturbations
are treated as quantum variables, while the background follows its classical behavior.
The goal of our approach is to consider also the quantum nature
of the background, and compute corrections to the dynamics derived within the QFT approximation.
In particular, we will be especially interested on the form of the power spectrum, for both
scalar and tensor modes, since it entails observable effects.

More precisely, we will consider the geometrodynamical framework for quantum gravity that leads
to the Wheeler--DeWitt equation. Therefore, starting from this equation,
the natural question is how to recover the equations of QFT on classical backgrounds plus
certain correction terms, which would encode the quantum-gravity effects.
An approach based on a Born--Oppenheimer type of decomposition of the wavefunction has been widely used in the literature
for inflationary scenarios \cite{kamenshchik_tronconi_venturi_20131_inflation_QG_BO}--\cite{CK21}. This approximation is based on a factorization of the wavefunction: one part only
depending on the {\it slow} degrees of freedom (background) and the other part encoding the information about the {\it fast}
degrees of freedom (perturbations) for a given configuration of the background.

Another alternative is to
perform a decomposition of the wavefunction into its infinite set of moments and follow the prescription for constrained
systems presented in \cite{bojowald_2009_effective_constraints_rel_Qsystems,bojowald_2008_effective_constraints_Qsystems}.
This formalism has been used to introduce a concept of time for semiclassical states in quantum cosmological models
\cite{BT10}--\cite{BH18}, and it has also already been applied to the analysis of the inflationary universe in \cite{brizuela_muniain_2019_moments_QG}.
In fact, the present paper will generalize the analysis performed in \cite{brizuela_muniain_2019_moments_QG}
in several important aspects. In particular, in that reference a truncation of the system at second-order
in moments was implemented. The corrections to the power spectrum were obtained to be quadratic in second-order
moments which, formally, would correspond to fourth-order terms. Therefore, in this paper we will explore
the third- and fourth-order truncation schemes in detail. In addition, it was assumed that at the onset of inflation
the system was on its fundamental state. Nevertheless, this might not be the case \cite{LPS97}--\cite{Broy2016}.
Hence, as already performed in the context of the Born--Oppenheimer approximation
\cite{Brizuela_2019_power_spectrum_excited_states}, here generically
excited initial states will be considered. Finally, the most important generalization is that
we will take into account the back-reaction produced by other modes on the evolution
of the mode under consideration. Up to our knowledge, this effect has been neglected in all
the previous studies in the literature, but our results will point out that it enters in
the leading-order correction term to the power spectrum and it is thus not negligible.

The rest of the paper is organized as follows. In Sec. \ref{sec:quantization} the quantization of
the system is performed. In Sec. \ref{section_moment_decomposition} the quantum moments are defined 
and the general method is explained. Sec. \ref{sec:third_order_truncation} considers a truncation
of third-order in moments and shows that, at this order, the power spectrum does get any correction term.
In Sec. \ref{sec:fourth-order}, after imposing a fourth-order truncation in the moments,
the power spectrum for initial excited states and under the presence of infinite modes is obtained.
Finally, our conclusions are presented in Sec. \ref{section_conclusions}. 

\section{Canonical quantization of an inflationary model}\label{sec:quantization}

We will consider an inflationary model given by a spatially flat
Friedmann-Lemaitre-Robertson-Walker metric minimally coupled to an inflaton field $\phi$ with potential $V(\phi)$.
As it is well known, by performing a harmonic decomposition on the homogeneous
spatial sections, the linear perturbations of this model can be classified
into three different types: scalar, vector and tensor perturbations.
At linear level different sectors evolve independently and thus can be treated separately.

By construction, tensor perturbations are invariant
under gauge transformations, while scalar and vector perturbations are not.
Nonetheless, one can perform suitable canonical transformations and
factor out all the gauge and constrained degrees of freedom, in such
a way that the complete physical information of the problem is contained
in three variables: on the one hand, the Mukhanov-Sasaki variable $v^s$,
which encodes the gauge-invariant perturbations of the inflaton field $\phi$.
And, on the other hand, $v^{+}$ and $v^{\times}$, which describe
the two polarizations of the gravitational wave. Let us, for compactness,
denote all these variables as $v^\sigma$, with $\sigma=s,+,\times$.

The smeared Hamiltonian constraint that describes the full (both background and perturbative) dynamics of this system
takes the following form,
\begin{equation}\label{hamiltonian0}
\int {\rm d}\eta\, {\mathcal H}:=\int {\rm d}\eta \int {\rm d}{\bf x}\, N\,\mathcal{C}=\int {\rm d}\eta\int {\rm d}{\bf x} \left[-\frac{G}{2}\moma^{2}+\frac{\momphi^{2}}{2a^{2}}+a^{4}V(\phi)\right]+\int {\rm d}\eta \,  H_{pert},
\end{equation}
where the lapse has been chosen equal to the
scale factor $N=a$. In addition, the reduced gravitational constant $G$ has been defined as $G:=\frac{4\pi G_N}{3}$, with $G_N$ being Newton's gravitational constant.
The variables $\pi_a:=-a' /{G}$ and $\pi_\phi:=a^2\phi'$
are, respectively, the conjugate momentum of the scale factor $a$
and of the inflaton field $\phi$, where the prime stands for derivative
with respect to the conformal time $\eta$.

Since we are assuming flat spatial slices,
in order to write down the perturbative Hamiltonian $H_{pert}$,
it is convenient to perform a Fourier transformation of the variables $v^{\sigma}$,
and their corresponding conjugate momenta $\pi^{\sigma}$, as follows,
\begin{eqnarray}
v^{\sigma}(\eta, {\bf x}) =\frac{1}{(2 \pi)^{3/2}}\int_{\mathbb{R}^3}{\rm d}{\bf k}\, u^\sigma_{\bf k}(\eta) e^{i  {\bf k \cdot x}},\\
  \pi^{\sigma}(\eta, {\bf x}) =\frac{1}{(2 \pi)^{3/2}}\int_{\mathbb{R}^3}{\rm d}{\bf k}\, p^\sigma_{\bf k}(\eta) e^{i  {\bf k \cdot x}}.
\end{eqnarray}
These complex Fourier modes obey the reality condition $u^\sigma_{-\bf k}=(u^\sigma_{\bf k})^*$ and $p^\sigma_{-\bf k}=(p^\sigma_{\bf k})^*$,
and the conjugate momentum of $u^\sigma_{\bf k}$ is $p^\sigma_{\bf -k}$,
that is, $\{u^\sigma_{\bf k},p^\sigma_{\bf q}\}=\delta({\bf k+q})$.
Nonetheless, in order to work with just real modes,
it is possible to perform a subsequent transformation
by simply decomposing each mode on its real and imaginary parts:
\begin{eqnarray}
u^\sigma_{\bf k}=\frac{1}{\sqrt{2}}(u^{\sigma, R}_{\bf k}+ i u^{\sigma,I}_{\bf k}), \\
p^\sigma_{\bf k}=\frac{1}{\sqrt{2}}(p^{\sigma, R}_{\bf k}+ i p^{\sigma,I}_{\bf k}),
\end{eqnarray}
in such a way that $\{u_{\bf k}^{\sigma,R},p_{\bf q}^{\sigma,R}\}=\delta(\bf k-q)$
and $\{u_{\bf k}^{\sigma,I},p_{\bf q}^{\sigma,I}\}=\delta(\bf k-q)$.
In terms of these real modes, the contribution to the Hamiltonian constraint
from the perturbative variables can be written in the following compact
form,
\begin{equation}
 H_{pert}=\frac{1}{2}\sum_{r=R,I}\sum_{\sigma=s,+,\times} \int {\rm d}{\bf k}\left[(p^{\sigma,r}_{\bf k})^{2}
 +\omega_{{\bf k},\sigma}^{2}\,(u_{\bf k}^{\sigma,r})^{2}\right],
\end{equation}
where the integration is taken over half Fourier space, that is, $\mathbb R^2\times {\mathbb R}^+$. Note that, in this expression there are two sums: one in $r$
over the real $(R)$ and imaginary $(I)$ part of the modes, and another
one in $\sigma$ over the three different degrees of freedom
(the scalar sector and the two polarizations of the tensor sector).
Therefore, for each wavevector $\bf k$, this Hamiltonian is the linear combination of six parametric
oscillators. In the classical treatment, as well as in the approximation given by QFT on
classical backgrounds, these six oscillators evolve independently.

Concerning the frequency $\omega_{{\bf k},\sigma}$,
it depends on the conformal time $\eta$, on the wavenumber $k$, and on the sector:
for the scalar modes it is given by $\omega_{{\bf k},s}^2:=k^2-a''/a$,
whereas for tensor modes it reads as $\omega_{{\bf k},+}^2=\omega_{{\bf k},\times}^2
:=k^2-z''/z$, with $z:=(2a'^2-a a'')^{1/2}a/a'=a(1-(a'/a)'/(a'/a)^2)^{1/2}$.
In consequence, since they obey the same equations of motion, the two oscillators corresponding
to the scalar sector will follow an identical evolution, as long as one chooses the same
initial state for both. This argument certainly also applies to the four oscillators of the tensor sector.
Even if in our analysis the full Hamiltonian
constraint \eqref{hamiltonian0} will be quantized and all these oscillators
will couple, we will also consider the same initial state for
the oscillators of a given sector, as it is done in QFT. Hence, one can perform the following transformation,
\begin{eqnarray}
v_{\bf k}^s &:=&\sqrt{2}u_{\bf k}^{s,R}=\sqrt{2}u_{\bf k}^{s,I},\\
\pi_{\bf k}^s&:=&\sqrt{2}p_{\bf k}^{s,R}=\sqrt{2}p_{\bf k}^{s,I},\\
v_{\bf k}^t&:=&2u_{\bf k}^{+,R}=2u_{\bf k}^{+,I}=2u_{\bf k}^{\times,R}=2u_{\bf k}^{\times,I},\\
\pi_{\bf k}^t&:=&2p_{\bf k}^{+,R}=2p_{\bf k}^{+,I}=2p_{\bf k}^{\times,R}=2p_{\bf k}^{\times,I},
\end{eqnarray}
that leaves just two harmonic oscillators per wavevector:
one for the scalar sector $(v_{\bf k}^s, \pi_{\bf k}^s)$ and one for the tensor sector $(v_{\bf k}^t, \pi_{\bf k}^t)$, with canonical brackets
$\{v_{\bf k}^\sigma,\pi_{\bf q}^{\lambda}\}=\delta({\bf k}-{\bf q})\delta_{\sigma\lambda}$.
These are the variables that will be considered for the rest of the analysis.

Let us now comment on the explicit spatial integral on the right-hand side of expression \eqref{hamiltonian0}.
Note that, due to homogeneity, none of the variables inside this integral
have any spatial dependence. Therefore, it is immediate to perform the integral,
which provides the spatial volume $L^3$ as a global factor. This volume can be absorbed by performing
the following rescaling of the scale factor $a$
and of the conformal time $\eta$:
\begin{equation}
 a\rightarrow a/L,\qquad \eta\rightarrow\eta L.
\end{equation}
The rescaling of the time $\eta$ implies that the momenta are being
rescaled as $\pi_a\rightarrow\pi_a/L^2$ and $\pi_\phi\rightarrow\pi_\phi/L^3$.
If the volume $L^3$ is finite, all the above integrals
in the Fourier space will be replaced by discrete sums. In particular, if one assumes periodic boundary conditions, in Cartesian
coordinates, the wavevector would be given by ${\bf k}=2\pi/L(n_1,n_2,n_3)$, and the integral
$\int {\rm d}{\bf k}$ would be replaced by the sum
$$\left(\frac{2\pi}{L}\right)^3\sum_{n_1, n_2,n_3},$$
running over half the Fourier space, which is
given by the following set of integers:
$\{\forall\, n_1,\forall\, n_2,n_3>0\}\cup\{\forall\, n_1,n_2>0,n_3=0\}\cup\{n_1>0,n_2=0,n_3=0\}$.
Furthermore, the discretization factor that appears in front of the sum can
be absorbed in the perturbative variables by a rescaling
$v_{\bf k}\rightarrow L^2/(2\pi)^{3/2} v_{\bf k}$ and $k\rightarrow k/L$. This rescaling, in combination with the above rescaling for
the time variable, implies a transformation $\pi_{\bf k}\rightarrow L/(2\pi)^{3/2} \pi_{\bf k}$ for the momenta.
Finally, the smeared Hamiltonian constraint takes the following form,
\begin{equation}\label{hamiltonian_classic_total_constraint}
\mathcal{H}=-\frac{G}{2}\moma^{2}+\frac{\momphi^{2}}{2a^{2}}+a^{4}V(\phi)+
\frac{1}{2}\sum_{\sigma=S, T}\sum_{n_1,n_2,n_3}(\pi_{\bf k}^{\sigma})^{2}+\omega_{{\bf k},\sigma}^{2}\,(v_{\bf k}^{\sigma})^{2},
\end{equation}
with canonical Poisson brackets $\{a,\pi_a\}=\{\phi,\pi_\phi\}=1$
and $\{v_{\bf k}^\sigma,\pi_{\bf q}^\lambda\}=\delta_{ \bf kq}\delta_{\sigma\lambda}$,
and ${\bf k}$ being given in terms of the integers $n_1$, $n_2$ and $n_3$, as commented
above. From this point on,
and in order to alleviate the notation, the sum that appears in the last
expression will be compactly denoted as $\sum_{{\bf k},\sigma}$. In addition,
since it will be possible to treat all the sectors on the same footing,
the $\sigma$ superindex will also be removed from the explicit notation of the different variables.
In this way, we will write the above Hamiltonian as,
\begin{equation}\label{class_ham}
\mathcal{H}=-\frac{G}{2}\moma^{2}+\frac{\momphi^{2}}{2a^{2}}+a^{4}V(\phi)+
\frac{1}{2}\sum_{{\bf k},\sigma}\pi_{\bf k}^{2}+\omega_{\bf k}^{2}\,v_{\bf k}^{2}.
\end{equation}

The canonical quantization of the system is then performed by promoting different
variables to operators, and Poisson brackets to commutators: $\{\cdot ,\cdot \}\rightarrow i\hbar[\cdot,\cdot]$.
Physical states are defined as those annihilated by the Hamiltonian operator,
which leads to the Wheeler--DeWitt equation,
\begin{equation}\label{WdWeq}
 \widehat{\mathcal H}\Psi=0.
\end{equation}
The goal of this paper is to solve this equation by
performing a decomposition of the wavefunction in its infinite
set of quantum moments. In particular, we will be interested on the
evolution of the fluctuation of the perturbative variable $v_{\bf k}$,
that defines the power spectrum. More precisely,
we will define the dimensionless quantity,
\begin{equation}
	\label{power_spectrum_general_adim}
 {\cal P}_{\bf k}:=\frac{G k^3}{a^2}\langle\, \widehat v_{\bf k}^2\rangle,
\end{equation}
which, up to global numerical factors, provides the power spectrum
for each sector.

\section{Moment decomposition}\label{section_moment_decomposition}

Let us define the central moments of the wavefunction as follows,
\begin{align}
\label{q_moments_modes}
&\Delta(a^{i_1}\moma^{i_2} \phi^{i_3}\momphi^{i_4}v_{\bf k_1}^{m_1}\dots v_{\bf k_n}^{m_n}\pi_{\bf q_1}^{n_1}\dots\pi_{\bf q_r}^{n_r}):=\\\nonumber
&\!\average{\fluct{a}^{i_1}
(\widehat{\pi}_a-\pi_a)^{i_2}
\fluct{\phi}^{i_3}
(\widehat{\pi}_\phi-\pi_\phi)^{i_4}
(\widehat{v}_{\bf k_1}-v_{\bf k_1})^{m_1}
\!\!\dots\!(\widehat{v}_{\bf k_n}-v_{\bf k_n})^{m_n}
(\widehat{\pi}_{\bf q_1}-\pi_{\bf q_1})^{n_1}
\!\!\dots\!
(\widehat{\pi}_{\bf q_r}-\pi_{\bf q_r})^{n_r}
}_{Weyl},
\end{align} 
where the subscript \emph{Weyl} stands for totally symmetric ordering, and the symbols
without hat stand for
the expectation value of their corresponding operator, that is, $X:=\average{\widehat X}$. 
The sum of the different powers on the above definition
($i_1+i_2+i_3+i_4+m_1+\dots+m_n+n_1+\dots+n_l$) will be referred as the order of the
quantum moment.

These moments form an infinite set of variables and encode the complete physical information
of the wavefunction. They are purely quantum variables and thus, in the classical limit, all of them are vanishing.
On the one hand, in this formalism, instead of solving the corresponding equation for the wavefunction, which
for the present model would be the Wheeler--DeWitt equation \eqref{WdWeq}, one deals directly with the
equations of motion for the moments. In addition, the coupling between the moments
and the classical equations of motion provide a direct measure of the quantum back-reaction.
On the other hand, the main disadvantage of this formalism is that
one deals with infinite variables. Therefore, in practice, a reduction of the system is usually necessary, which
can be achieved by simply dropping moments of an order higher
than a given truncation order. This kind of truncation is valid for peaked semiclassical states,
since for such states a moment of order $n$ has a value of the order of ${\cal O}(\hbar^{n/2})$.

In a previous study \cite{brizuela_muniain_2019_moments_QG}, this system was analyzed with a
truncation at second order. Here, we will study the third and fourth-order truncations, in order to check how
these higher-order moments affect the physical results. Nonetheless, as already commented above,
this is not the only generalization of the mentioned study. We are also considering
the presence of infinite perturbative modes, which means that,
even with a truncation at a given order in moments,
we will still be dealing with an infinite number of variables.

In order to convert the Wheeler--DeWitt equation \eqref{WdWeq} on a system
of equations for the moments, we will follow the method presented in
\cite{bojowald_2009_effective_constraints_rel_Qsystems}.
The main idea of this method is that, since
the action of the Hamiltonian $\widehat{\cal H}$ on any physical wavefunction
must be zero, the expectation value of the Hamiltonian $\langle\widehat{\cal H}\rangle$,
as well as the expectation value of the Hamiltonian multiplied from the left
by any operator $\langle\widehat O\widehat{\cal H}\rangle$, should also
be vanishing.
By adequately choosing the operator $\widehat O$ as the
symmetrically ordered product of basic fluctuations,
\begin{align}
\widehat O=\bigg[
\fluct{a}^{i_1}(\momaop-\moma)^{i_2}\fluct{\phi}^{i_3}(\momphiop-\momphi)^{i_4}(\widehat{\pi}_{\bf k_1}-\pi_{\bf k_1})^{m_1}\!\!\dots\!(\widehat{\pi}_{\bf k_n}-\pi_{\bf k_n})^{m_n}
(\widehat{\pi}_{\bf q_1}-\pi_{\bf q_1})^{n_1}\!\!\dots
\nonumber\\
\dots\!
(\widehat{\pi}_{\bf q_r}-\pi_{\bf q_r})^{n_r}\bigg]_{Weyl},
\end{align}
and writing the mentioned expectation values in terms of the moments, one obtains
a system of first-class relations that constraint the moments.
These constraints encode a gauge freedom of the system, and one
can then fix the gauge following the usual prescription for constrained
systems. In the present context, this gauge freedom is related to
the choice of a time variable. Classically there is just one constraint
$({\cal H}=0)$, which allows one to choose one time variable and deparametrize
the Hamiltonian. The new quantum constraints that appear at every order
in moments are related
to the choice of fluctuations and higher-order moments of the time variable.

In order to choose an internal time variable, the natural gauge-fixing conditions would be that
all the moments related to that variable should be vanishing; so that
it can be understood as a parameter, instead of a physical
degree of freedom. In our case, as performed in the previous analysis \cite{brizuela_muniain_2019_moments_QG},
the scale factor $a$ will be chosen as the internal time variable,
which is a monotonically increasing function. In particular,
this is a well-behaved time variable in the case of a constant potential
that will be discussed later; unlike the matter variables $(\phi,\pi_\phi)$, which are constant in that case. 
The system of constraints will be solved for its conjugate momentum
$\pi_a$, which will then become the physical Hamiltonian of the deparametrized system.

\section{Third-order truncation}\label{sec:third_order_truncation}

In this section we will consider a third-order truncation in moments.
In the first subsection we will assume a generic potential $V(\phi)$ for the inflaton field.
The full system of first-class constraints will be obtained and, by choosing the scale factor $a$
as the time parameter, it will be solved in order to obtain the physical Hamiltonian of the system
${\cal H}_a:=-\pi_a$. In the second subsection the dynamics for the constant-potential
case will be analyzed in detail.

\subsection{Deparametrization of the model for a general potential}

At third order the relevant constraint equations for the moments will be given by the following expectation values:
\begin{eqnarray}
\average{\widehat{\cal H}}&=&0,\label{exph}\\
\average{\fluct{X}\widehat{\cal H}}&=&0, \label{expxh}\\
\label{expxyh}
\frac{1}{2}\average{\left[\fluct{X}\fluct{Y}+\fluct{Y}\fluct{X}\right]\widehat{\cal H}}&=&0, 
\end{eqnarray}
where $X$ and $Y$ stand for all our basic variables: $(a, \pi_a, \phi, \pi_\phi, v_{\bf k},\pi_{\bf k})$.
By performing an expansion around the expectation values, equation \eqref{exph}
can be written in terms of the moments as follows,
\begin{eqnarray*}
 \average{\widehat{\cal H}}&=&
\sum \frac{\partial^{N}\mathcal{H}}
{\partial a^{i_1}\partial\moma^{i_2} \partial\phi^{i_3}\partial\momphi^{i_4}\partial v_{\bf k_1}^{m_1}\dots \partial v_{\bf k_n}^{m_n}\partial \pi_{\bf q_1}^{n_1}\dots\partial \pi_{\bf q_r}^{n_r}}
\frac{ \Delta(a^{i_1}\moma^{i_2} \phi^{i_3}\momphi^{i_4}v_{\bf k_1}^{m_1}\dots v_{\bf k_n}^{m_n}\pi_{\bf q_1}^{n_1}\dots\pi_{\bf q_r}^{n_r})}
{i_1!i_2!i_3!i_4!m_1!\dots m_n!n_1!\dots n_r!}=0,
\end{eqnarray*}
where $\mathcal{H}$ is the classical Hamiltonian \eqref{class_ham}, $N$ is the order of the derivative $(N:=i_1+i_2+i_3+i_4+m_1+\dots+ m_n+n_1+\dots+n_r)$ and
the sum runs over all nonnegative integer values for all the indices $i_j, m_j$ and $n_j$.
In addition, in this expression $\Delta(1)=1$ and $\partial^0 {\cal H}={\cal H}$
should be understood, and $\Delta(X)=0$ by definition. 
If one truncates this expression at third order in moments,
the expectation value of the Hamiltonian takes the following explicit form,
\begin{eqnarray}
\average{\widehat{\cal H}}&=&-\frac{G}{2}\moma^{2}+\frac{\momphi^{2}}{2a^{2}}+a^{4}V(\phi)+
\frac{1}{2}\sum_{{\bf k},\sigma}\left(\pi_{\bf k}^{2}+\omega_{\textbf{k}}^{2}\,v_{\bf k}^{2}\right)-\frac{G}{2}\Delta( \moma^2)
+\frac{3 \pi _{\phi }^2 \Delta (a^2)}{2a^4}+\frac{3 \pi _{\phi }  \Delta(a^2\momphi)}{a^4}
\nonumber\\
&
-&\frac{2 \pi _{\phi } \Delta(a\momphi)}{a^3}
-\frac{2 \pi _{\phi }^2 \Delta( a^3)}{a^5}+\frac{\Delta(\momphi^2)}{2a^2}-\frac{\Delta(a\momphi^2)}{a^3}
+2 a\left[3 a \Delta(a^2) V(\phi ) +2 \Delta (a^3)\right] V(\phi )
\nonumber
\\
&+&2a^2 \left[ 2 a \Delta(a\phi) +3 \Delta(a^2\phi)\right] V'(\phi )+\frac{a^3}{2}\left[a \Delta(\phi^2)+4 a^6 \Delta(a\phi^2)\right] V''(\phi )+\frac{ a^4}{6} \Delta(\phi^3) V^{(3)}(\phi )
\nonumber
\\\label{constraint_classical_mode_3}
&+&\frac{1}{2}\sum_{{\bf k},\sigma}\left(\Delta(\pi_{\bf k}^{2})+\omega_{\textbf{k}}^{2}\,\Delta(v_{\bf k}^{2})\right)=0.
\end{eqnarray}
The first four terms correspond to the classical Hamiltonian \eqref{class_ham}, whereas the rest are
terms linear in moments that encode the quantum back-reaction of the state. In particular,
it is clear from this expression that, upon quantization, the classical constraint is no
longer fulfilled, as all the moments can not be vanishing at the same time.

Concerning the other two types of constraints that we need to consider, \eqref{expxh}--\eqref{expxyh},
it can be shown that at third order they take the following form,
\begin{eqnarray}
\label{linearconstraints}
&&\!\!\!\!\!\!\!\!\!\!\!\!\!\!\!\!\!\!\!\!\!\!\!\langle (\widehat{X} - X)\widehat{\cal H}\rangle =\frac{i\hbar}{2}\frac{\partial \mathcal{H}}{\partial p_X} \\\nonumber
&&+
\sum \frac{\partial^{N}\mathcal{H}}
{\partial a^{i_1}\partial\moma^{i_2} \partial\phi^{i_3}\partial\momphi^{i_4}\partial v_{\bf k_1}^{m_1}\dots \partial v_{\bf k_n}^{m_n}\partial \pi_{\bf q_1}^{n_1}\dots\partial \pi_{\bf q_r}^{n_r}}
\frac{ \Delta(Xa^{i_1}\moma^{i_2} \phi^{i_3}\momphi^{i_4}v_{\bf k_1}^{m_1}\dots v_{\bf k_n}^{m_n}\pi_{\bf q_1}^{n_1}\dots\pi_{\bf q_r}^{n_r})}
{i_1!i_2!i_3!i_4!m_1!\dots m_n!n_1!\dots n_r!}
,
\end{eqnarray}
with $p_X$ being the conjugate momentum of the variable $X$ under consideration, and
\begin{eqnarray}\label{quadraticconstraints}
&&\!\!\!\!\!\!\!\!\!\!\!\!\!\!\!\!\!\!\!\!\!\!\!\frac{1}{2}\langle [(\widehat{X} - X)(\widehat{Y} - Y)+(\widehat{Y} - Y)(\widehat{X} - X)]\widehat{\cal H}\rangle=\\\nonumber
&&\sum \frac{\partial^{N}\mathcal{H}}
{\partial a^{i_1}\partial\moma^{i_2} \partial\phi^{i_3}\partial\momphi^{i_4}\partial v_{\bf k_1}^{m_1}\dots \partial v_{\bf k_n}^{m_n}\partial \pi_{\bf q_1}^{n_1}\dots\partial \pi_{\bf q_r}^{n_r}}
\frac{ \Delta(XYa^{i_1}\moma^{i_2} \phi^{i_3}\momphi^{i_4}v_{\bf k_1}^{m_1}\dots v_{\bf k_n}^{m_n}\pi_{\bf q_1}^{n_1}\dots\pi_{\bf q_r}^{n_r})}
{i_1!i_2!i_3!i_4!m_1!\dots m_n!n_1!\dots n_r!}.
\end{eqnarray}
The $X$ (and $Y$) inside the $\Delta$ that defines the moment means that the order of the variable $X$ (and $Y$)
under consideration should be increased by one.
For instance, for the case $X=a$ and $Y=\phi$,
$\Delta(XYa^{i_1}\moma^{i_2} \phi^{i_3}\momphi^{i_4}v_{\bf k_1}^{m_1}\dots v_{\bf k_n}^{m_n}\pi_{\bf q_1}^{n_1}\dots\pi_{\bf q_r}^{n_r}):=\Delta(a^{i_1+1}\moma^{i_2} \phi^{i_3+1}\momphi^{i_4}v_{\bf k_1}^{m_1}\dots v_{\bf k_n}^{m_n}\pi_{\bf q_1}^{n_1}\dots\pi_{\bf q_r}^{n_r})$.
In principle, in these last two expressions the sum should run over all nonnegative integers $i_j$, $m_j$ and $n_j$.
But, since we are considering a truncation at third order in moments, the order of the derivative
$N$ (defined again as $N:=i_1+i_2+i_3+i_4+m_1+\dots+ m_n+n_1+\dots+n_r$) would just take the values $N=0,1,2$
in \eqref{linearconstraints}, whereas in the expression \eqref{quadraticconstraints} it has only two possible values $N=0,1$.
For illustration purposes,
the constraints \eqref{linearconstraints} are given in the App. \ref{appendix_constraints} for
each basic variable $X$, but
the constraints \eqref{quadraticconstraints} are quite lengthy and we will refrain from displaying them explicitly.

Note that, as one would expect, the constraints \eqref{linearconstraints} are complex. The imaginary
term comes from the reordering of the operators and, at this order, is given by the derivative of the
classical Hamiltonian constraint ${\cal H}$ with respect to the
conjugate momentum of the variable under consideration. Nonetheless, the constraints \eqref{quadraticconstraints}
are real. In this latter case, the reordering terms are of order $\hbar^2$, which corresponds to fourth order in moments,
and thus have been dropped for the present third-order approximation.

As already commented above,
these constraint equations are related to a gauge freedom. Hence, following the standard procedure, one needs to fix a gauge
condition which, in this context, amounts to the choice of a time variable. In our case, the most natural time variable is the scale
factor, as it is a monotonically increasing function during an inflationary
evolution. Other expectation values, like $\pi_a$, $\pi_\phi$,
$v_{\bf k}$ and $\pi_{\bf k}$, are vanishing (or very small),
whereas the scalar field $\phi$ is (approximately) constant during the slow-roll evolution.
Choosing $a$ as the time parameter implies that all the moments related to $a$,
that is, of the form $\Delta(Xa)$ and  $\Delta(XYa)$, for all $X, Y\neq \pi_a$, should be chosen to be vanishing.
Thus, these will be our gauge-fixing conditions. The constraints \eqref{constraint_classical_mode_3}--\eqref{quadraticconstraints} should then be understood as equations to be solved
for the moments related to $\pi_a$ (that is, $\Delta(X\pi_a)$ and $\Delta(XY\pi_a)$ for all $X, Y$), as well as $\pi_a$ itself.
This procedure will deparametrize the system and define ${\cal H}_a:=-\pi_a$ as our physical Hamiltonian.

At this point, it is very instructive to count degrees of freedom and constraint equations.
For definiteness let us, just for this purpose, assume the presence of only two different perturbative modes: ($v_{\bf k}$, $\pi_{\bf k}$) and  ($v_{\bf q}$, $\pi_{\bf q}$).
In this way, one would have eight independent basic variables ($a$, $\pi_a$, $\phi$, $\pi_\phi$, $v_{\bf k}$, $\pi_{\bf k}$, $v_{\bf q}$, $\pi_{\bf q}$),
and thus one would get eight constraints of the form \eqref{linearconstraints}.
Concerning constraints of the form \eqref{quadraticconstraints}, there would be 36 of them, one for each independent
couple of the basic variables. Even if large,
the system of equations \eqref{linearconstraints}--\eqref{quadraticconstraints}, is linear in moments.
Following the commented gauge-fixing procedure, these 44 equations should be solved for the 44 moments related to
$\pi_a$; that is, the 8 second-order moments of the form $\Delta(X\pi_a)$ and the 36 third-order moments
of the form $\Delta(XY\pi_a)$, for all $X, Y$. The last step would then be to replace this solution in the Hamiltonian
constraint \eqref{constraint_classical_mode_3}, and solve it for $\pi_a$. In this way, and requesting $\Delta(Xa)=0=\Delta(XYa)$,
for all $X,Y\neq\pi_a$, one would get the form of the physical Hamiltonian ${\cal H}_a:=-\pi_a$ in terms of
the time $a$, the physical variables ($\phi$, $\pi_\phi$, $v_{\bf k}$, $\pi_{\bf k}$, $v_{\bf q}$, $\pi_{\bf q}$),
and their moments $\Delta(XY)$ and $\Delta(XYZ)$, with $X,Y,Z\neq a,\pi_a$.

Note that the solution to the linear system \eqref{linearconstraints}--\eqref{quadraticconstraints}
gives the moments $\Delta(X\pi_a)$ and $\Delta(XY\pi_a)$ in terms of moments unrelated to $\pi_a$
with coefficients that depend on expectation values, and in particular on powers of $\pi_a$. In
general, the higher the truncation under consideration, the higher the powers of $\pi_a$ that would
appear in those coefficients as one considers expectation values of higher powers of operators.
Therefore, when
applying the commented procedure under the presence of infinite modes, and using the solution
to the linear system \eqref{linearconstraints}--\eqref{quadraticconstraints} to remove moments
of the form $\Delta(X\pi_a)$ and $\Delta(XY\pi_a)$, the Hamiltonian constraint
\eqref{constraint_classical_mode_3} turns into the following sixth-order polynomial equation for $\pi_a$:
\begin{equation}
\label{eqmoma_modes_third}
G^3\moma^{6}-G^2\left(G p_a^2+A_{4}\right)\moma^{4}+ i\hbar G^2 A_{3}\moma^{3}
+G A_{2}\moma^{2}
+A_{0}=0,
\end{equation}
where $p_a$ is defined as the classical value of $\pi_a$,
\begin{equation}\label{padef}
 p_a:=-\frac{1}{\sqrt{G}}\left[\frac{\pi_\phi^2}{a^2}+2a^4 V(\phi)+\sum_{{\bf k},\sigma} \pi_{\bf k}^2+\omega_k^2 v_{\bf k}^2 \right]^{1/2},
\end{equation}
and the coefficients $A_{i}$ are real functions of the expectation values and moments unrelated to $\moma$.
In fact, the coefficients $A_3$ and $A_4$ have a quite simple form,
\begin{align}
	A_3&=\frac{\momphi^2}{a^3}-4a^3V(\phi),\\
	A_4&=\frac{19}{16} \deltados{\momphi}+\frac{19}{16}  a^4 \deltados{\phi}V''(\phi )+\frac{1}{3}  a^4 \deltatres{\phi} V^{(3)}(\phi )
+\frac{19}{16}  \sum_{\mathbf{k},\sigma}\left(\omega_{\mathbf{k}}^2 \deltados{\vksigma}+\deltados{\momvsigma}\right),
\end{align}
whereas $A_0$ and $A_2$ are much lengthier and are given in App. \ref{appendix_coefficients_A0_A2_third_order}.
Note that
the only imaginary term that appears in equation \eqref{eqmoma_modes_third} is multiplied by the coefficient $A_3$,
which is equal to the time derivative of the Hamiltonian $\partial {\cal H}/\partial a$.

We are interested in the (slow-roll) inflationary evolution of the system. And, even if we are considering
quantum corrections, this formalism is valid as long as the evolution is close to a classical trajectory.
Hence, let us study how the different coefficients behave during such an evolution.
In particular, the scale factor goes as $a\approx e^{Ht}$, with $t$ being the cosmological time and $H$
the (approximately) constant Hubble factor. The rest of the variables behave as $\moma\approx e^{2Ht}$, $\momphi=0$, $V(\phi)\approx H^2$, $\vj\approx 0$, and $\momvj\approx 0$.
Using these dependencies, we find that all the terms in the equation \eqref{eqmoma_modes_third}, except the imaginary term, increase as follows
$$\moma^{6}\approx A_{4}\moma^{4} \approx A_2 \moma^2\approx A_0 \approx e^{12Ht}.$$
The imaginary term also grows with time but much slower, concretely as $ A_{3}\moma^{3}\approx e^{9Ht}$,
and thus can be neglected in relation with the rest. Therefore, under this assumption, equation \eqref{eqmoma_modes_third} is simplified to the following form
\begin{equation}
 \label{eqmoma_modes}
G^3\moma^{6}-G^2\left(G p_a^2+A_{4}\right)\moma^{4}
+G A_{2}\moma^{2}
+A_{0}=0.
\end{equation}
This is a third-order polynomial equation for $\moma^2$, with three independent solutions.
Out of these three solutions, two happen to be complex. Therefore our physical Hamiltonian
will be given by the unique real solution.
Even if it has a quite complicate form, taking into account that $A_0$, $A_2$ and $A_4$ are linear in
second- and third-order moments, we can make a Taylor expansion of the solution in the order of the moments,
and write the physical Hamiltonian as
\begin{equation}
	\label{hamiltonian_third_order_ai}
 {\cal H}_a:=-\pi_a=-p_a+\frac{A_0}{2 G^3 p_a^5}+\frac{A_2}{2 G^2
   p_a^3}-\frac{A_4}{2 G p_a}.
\end{equation}
From this expression one can clearly see that in the classical limit one would
recover $\pi_a=p_a$, since all the coefficients $A_0$, $A_2$, $A_4$ would vanish.
Note also that the sign of $p_a$ \eqref{padef} has been chosen as negative in order to get
a positive Hamiltonian.

This last expression is the main result of this subsection, and provides a physical Hamiltonian
that rules the quantum evolution of this model for any potential $V(\phi)$, with the scale factor
$a$ playing the role of the evolution parameter. Due to this choice, there are no quantum moments
related to the degree of freedom $(a,\pi_a)$: moments related to $a$ are directly fixed by the
gauge-fixing condition, whereas moments related to its conjugate momentum $\pi_a$ are determined
by solving the constraint equations. Therefore, the physical variables will be the expectation
values $(\phi, \momphi, \vk,\momv)$, along with its fluctuations and the correlations between them,
as well as their third-order moments. The evolution equations for different variables can be obtained
by computing their Poisson brackets with this Hamiltonian, and making use of the usual relation that
provides the Poisson brackets between expectation values in terms of the commutator between the corresponding operators:
$\{\langle \widehat A\rangle,\langle \widehat B\rangle\}=-i/\hbar\langle[\widehat A,\widehat B]\rangle$.

\subsection{Constant potential}\label{subsection_de_sitter_universe}

Let us now analyze the dynamics of the system for the particular, but relevant, case of a constant
potential $V(\phi)=H^{2}/(2G)$, with $H$ being the constant Hubble factor.
At a classical level this model corresponds to the de Sitter universe, and it provides the dominant contribution to any slow-roll
inflationary dynamics. For this case, the frequency of the modes coincides both for the tensor and scalar sectors, and takes
the simple form $\omega^{2}_{\bf k}=k^{2}-2a^{2}H^{2}$.

Classically, if one chooses an initially vanishing value for the momentum $\pi_\phi$, the inflaton field $\phi$ is constant
and $\pi_\phi$ vanishes all along evolution.
Nonetheless, this does not need to be the case for the quantum evolution, since the back-reaction terms
might produce a nonvanishing time derivative of either of these variables. But we will show below that,
by choosing an adequate initial (adiabatic) state at the onset of inflation,
this will also be obeyed by the quantum evolution.

At the beginning of inflation
($a\rightarrow 0$), the mode is well inside the horizon and the frequency tends to
the wavenumber $(\omega_{\bf k}\rightarrow k)$. Therefore, in the context of QFT
on classical backgrounds, one usually considers the stationary state of
a free mode on a flat Minkowski background.
In particular, this implies the vanishing of the expectation value of the perturbative variable
and its conjugate momentum: $v_{\bf k}=\pi_{\bf k}=0$.
Let us then analyze whether it is possible in our model to choose an adiabatic initial state, that
evolves coherently and in particular
keeps the initial conditions $v_{\bf k}=\pi_{\bf k}=\pi_\phi=0$, as well as the value of the inflaton $\phi$,
constant. Under these conditions, the evolution of these four variables would be given by:
\begin{eqnarray}
\label{v_evolution_equation}
\frac{d\vk}{da}&=&\{\vk,\ham\}=-\frac{G}{2a^6H^3}\left[\frac{\deltadosuno{\momphi}{\momv}}{a^2}+\sum_{\mathbf{q},\sigma}\Delta(\pi_{\bf k}\pi_{\bf q}^2)+\omega_{\mathbf{q}}^2
\Delta(\pi_{\bf k} v_{\bf q}^2)\right],\\
\label{momv_evolution_equation}
\frac{d\momv}{da}&=&\{\momv,\ham\}=
\frac{G\omega_{\mathbf{k}}^2}{2a^6H^3}\left[\frac{\deltadosuno{\momphi}{\vk}}{a^2}+\sum_{\mathbf{q},\sigma}\omega_{\mathbf{q}}^2 \Delta(v_{\bf k}v_{\bf q}^2)+\Delta(v_{\bf k}\pi_{\bf q}^2)\right],\\
\label{phi_evolution_equation}
\frac{d\phi}{da}&=&\{\phi,\ham\}=
-\frac{G}{2a^8H^3}\left[\frac{\deltatres{\momphi}}{a^2}+\sum_{\mathbf{q},\sigma}\left(\omega_{\mathbf{q}}^2 \Delta(\pi_\phi v_{\bf q}^2)+\Delta(\pi_\phi\pi_{\bf q}^2)\right)\right],\\
\label{momphi_evolution_equation}
\frac{d\momphi}{da}&=&\{\momphi,\ham\}=0.
\end{eqnarray}
The form of these equations suggest that one should choose an initial state with,
\begin{eqnarray}\nonumber
\Delta(\pi_{\bf k} v_{\bf q}^2)=\Delta(\pi_{\bf k}\pi_{\bf q}^2)=\deltadosuno{\momphi}{\momv}=0,\\
 \deltaunodos{\vk}{\momvi}=\Delta(v_{\bf k}v_{\bf q}^2)=\deltadosuno{\momphi}{\vk}=0,\nonumber\\\label{0moments1}
 \deltaunodos{\momphi}{\vk}=\deltaunodos{\momphi}{\momv}=\deltatres{\momphi}=0,
 \end{eqnarray}
so that the right-hand side of the above equations of motion is vanishing. The next step,
in order to construct an adiabatic state, is to compute the evolution of these set of moments and
check whether one can choose an initial state so that their time derivative is vanishing.
This is in fact the case, under the condition that the following moments are also
vanishing,
\begin{align}
	\deltauno{\vk\vi\momvj}=\deltauno{\vk\vi\vj}&=0,\nonumber\\
\deltauno{\vk\momvi\momvj}=\deltauno{\momv\momvi\momvj}&=0,\nonumber\\\label{0moments2}
\deltauno{\momphi\vk\vj}=\deltauno{\momphi\momv\momvi}&=0.
\end{align}
Under all the above conditions, the time derivative of this last set of moments also vanish,
and makes the construction of the initial adiabatic state consistent. Therefore, in summary, if one chooses an initial
state with vanishing value of the expectation values $v_{\bf k}$,$\pi_{\bf k}$, $\pi_\phi$ and of
the third-order moments \eqref{0moments1}--\eqref{0moments2}, all these variables will be vanishing
all along the evolution and $\phi$ will be a constant of motion.
(Note that there are certain third-order moments, as those related to the scalar field
$\phi$, which would not be vanishing.)
For such a state, the Hamiltonian \eqref{hamiltonian_third_order_ai} takes the following simple form:
\begin{align}
\label{physical_hamiltonian_modes}
\mathcal{H}_{a}&= \frac{a^2H}{G}
+\frac{(\Delta \momphi)^2}{2a^4H}
+\frac{1}{2a^2H}\sum_{\mathbf{k},\sigma}\left(\deltados{\momvsigma}+\omega^2_{\bf k} \deltados{\vksigma}\right).
\end{align}

We are particularly interested in computing the power spectrum of the perturbations,
which is encoded in the fluctuation $\deltados{\vk}$. In the present approximation,
the evolution equation of this fluctuation is only explicitly coupled to the fluctuation of the momentum $\deltados{\momv}$
and the correlation $\deltauno{\vk\momv}$. The equations for these three variables read as follows,

\begin{align}
\label{evolution_modes_vuno}
\frac{d}{da}(\deltauno{\vk\momv})&=\frac{1}{a^2H}\left[\deltados{\momv}-\omega_{\bf k}^2 \deltados{\vk}\right],\\
\label{evolution_modes_vunomomvuno}
\frac{d}{da}(\deltados{\vk})&=\frac{2}{a^2 H}\deltauno{\vk\momv},\\
\label{evolution_modes_momvuno}
\frac{d}{da}(\deltados{\momv})&=-\frac{2\omega_{\bf k}^2 }{a^2 H}\deltauno{\vk\momv}.
\end{align}
As it can be seen, there is no contribution from third-order moments, and in fact these equations are
the same as the ones obtained in the approximation of QFT on classical backgrounds. Therefore,
we conclude that the third-order terms will not contribute to modify the power spectrum. In the next section,
we will then move on to study the fourth-order truncation, which will indeed produce some
relevant contributions.

\section{Fourth-order truncation}\label{sec:fourth-order}

The analysis at fourth order would proceed in the same way as for third order:
one has to write the Wheeler--DeWitt equation as a set of constraints for the
moments, impose the gauge conditions ($\Delta(Xa)=0=\Delta(XYa)$,
for all $X,Y\neq\pi_a$), and solve them for $\pi_a$ and its moments. This procedure would define the
physical Hamiltonian ${\cal H}_a=-\pi_a$, which would then provide the evolution equations for the different
variables. Nonetheless, as will be explained below, at this order of truncation,
it will not be possible to complete this algorithm analytically and some assumptions
will have to be performed.

More precisely,
at fourth order, the set of constraints is given by the equations
we have already considered at the third-order truncation level, that is,
\eqref{exph}--\eqref{expxyh}, as well as the new set of constraints,
\begin{equation}\label{expxyzh}
 \langle (\widehat X-X)(\widehat Y-Y)(\widehat Z-Z)|_{Weyl} \widehat{\cal H} \rangle=0.
\end{equation}
Although this adds a large number of equations to the system,
these constraints, as well as the previous ones \eqref{expxh}--\eqref{expxyh}, are linear in moments.
Therefore, one can solve all these equations for moments related to
$\pi_a$, and replace their form in the Hamiltonian constraint \eqref{exph},
along with the gauge conditions $\Delta(XYa)=\Delta(Xa)=0$, for all $X,Y\neq\pi_a$.
In this way, one ends up with an equation, which does not contain
any moment related to the degree of freedom $(a,\pi_a)$, and that
should be solved for $\pi_a$ in order to deparametrize the system.
But here appears the main technical difficulty: whereas within
the third-order truncation this equation was the sixth-order polynomial equation \eqref{eqmoma_modes_third},
at fourth-order the resulting equation is not polynomial in $\pi_a$. In fact, it contains
complicated rational expressions of $\pi_a$ and it is thus not possible to obtain
a closed analytic form for its solution. There are two main contributions that produce such
complicated expressions. On the one hand, when considering higher-order truncations, one
needs to consider expectation values of higher powers of operators which, when expanded
in moments, introduce higher powers of the expectation values, and in particular of $\pi_a$, in the coefficients
of the moments. On the other hand, the coupling of the equations \eqref{expxh}--\eqref{expxyh} and \eqref{expxyzh}
is more involved than at previous orders, since a given moment appears in more constraints.
Therefore, linear combinations involving a larger number of equations, and with more complicated
coefficients of the expectation values, have to be considered to solve the system for each moment.

Hence, in order to analyze the behavior of the system with a fourth-order truncation in moments,
we will implement several simplifying assumptions. On the one hand, we will consider from the beginning the case
of a constant potential $V=H^2/(2 G)$, with the frequencies given by the expression $\omega_{\bf k}^2=k^2-2a^2H^2$.
On the other hand, we will assume that some properties of the system, that have
been shown to be valid up to third order, are still valid at the next level of truncation.
In particular, $\vk=\momv=\pi_\phi=0$ will be considered to be an exact solution of
the system. Since the Poisson brackets between any moment and the expectation values $(\vk,\momv,\pi_\phi)$
are vanishing, this solution can then be strongly imposed to construct the Hamiltonian for the moments.
Finally, we will also assume that
different imaginary terms do not contribute to the solution in a relevant way, so that
it is safe to neglect them. In this way, the Hamiltonian constraint
takes the simple form,
\begin{align}
	\label{constraint_classical_mode_4}
	\average{\widehat{\cal H}}&=-\frac{G}{2}\pi _a^2+\frac{a^4 H^2}{2G}+\frac{\Delta(\momphi^2)}{2a^2}-\frac{G}{2}\Delta( \moma^2)
	+\frac{1}{2}\sum_{{\bf k},\sigma}\left(\Delta(\pi_{\bf k}^{2})+\omega_{\bf k}^{2}\,\Delta(v_{\bf k}^{2})\right)=0,
\end{align}
where the gauge conditions $\Delta(Xa)=\Delta(XYa)=0$ have already been imposed.
The nontrivial part of this equation is contained in the fluctuation $\Delta(\moma^2)$,
which must be written in terms of moments unrelated to $\pi_a$ by solving the other
constraint equations \eqref{expxh}, \eqref{expxyh} and \eqref{expxyzh}. The expression that is obtained
for $\Delta(\moma^2)$ 
(as a function of $\pi_a$) is quite involve and, even under the present assumptions,
makes \eqref{constraint_classical_mode_4} to be
a quite complicate equation for $\pi_a$.
Thus, we will perform a Taylor expansion on the order of the moments,
and assume that the solution can be written as,
\begin{equation}
 \pi_a=p_a+p_2+p_3+p_4,
\end{equation}
with $p_a$ being the classical value of $\pi_a$ \eqref{padef}, and $p_2$, $p_3$ and $p_4$ being
respectively the contribution of second-, third- and fourth-order. One can then solve equation
\eqref{constraint_classical_mode_4}
iteratively order by order, which leads to the following form for the physical Hamiltonian,
\begin{align}
	\mathcal{H}_a:=-\pi_a&=\frac{a^2H}{G}+\frac{\deltados{\momphi}}{2 a^4 H}
	-\frac{G}{8 a^{10} H^3}\Delta(\momphi^4)+ \frac{\hbar^2G}{8a^6 H^3}\left(4a^2H^2+\sum_{{\bf k},\sigma}\omega_{\bf k}^2\right)
		\nonumber\\
		&+\frac{1}{2 a^2 H}\sum_{{\bf k},\sigma}\left(\deltados{\momv}+ \omega_{\bf k}^2\deltados{\vk} \right)
		-\frac{G}{4a^8H^3}\sum_{{\bf k},\sigma}\left( \Delta{(\momv^2\pi_\phi^2)}+ \omega_{\bf k}^2\Delta{(\vk^2\pi_\phi^2)}\right)
		\nonumber\\&
		-\frac{G}{8 a^6 H^3}\sum_{{\bf k},\sigma}\sum_{{\bf q},\sigma'}\left(\omega_{\mathbf{q}}^2\deltauno{\momv^2\vi^2}+\omega_{\mathbf{k}}^2\deltauno{\momvi^2v^2_{\bf k}}+\deltauno{\momv^2\momvi^2}+\omega_{\mathbf{k}}^2\omega_{\mathbf{q}}^2\deltauno{\vk^2\vi^2}\right).
\end{align}
As can be seen, unlike third-order moments, there are several fourth-order moments present
in this Hamiltonian, which in principle will back-react on the evolution of second-order moments.

At this point, in order to obtain the evolution equations for different moments, one just needs to
compute the Poisson brackets between each moment and this Hamiltonian. In particular,
we are interested in obtaining the power
spectrum \eqref{power_spectrum_general_adim}, which is directly related to the fluctuation of the perturbative variable $\deltados{\vk}$. The evolution equations for the three second-order moments $\Delta (\vk^2)$, $\Delta(\momv^2)$,
$\Delta(\vk\momv)$ take the form,
\begin{align}
	\label{evolution_equations_4_vk}
\frac{d}{d\xi}(\deltados{\vk})&=-\frac{2}{k}\deltauno{\vk\momv}+S_1,\\
\label{evolution_equations_4_vkmomv}
\frac{d}{d\xi}(\deltauno{\vk\momv})&=-\frac{1}{k}\left[\deltados{\momv}-\omega_{\bf k}^2 \deltados{\vk}\right]+S_2,\\
\label{evolution_equations_4_momv}
\frac{d}{d\xi}(\deltados{\momv})&=\frac{2}{k}\omega_{\bf k}^2 \deltauno{\vk\momv}+S_3,
\end{align}
where we have introduced the dimensionless time variable $\xi:=k/(aH)$ in order to see more clearly how the dependence
on the wavenumber $k$ enters the different equations.
Note that, except the source terms $S_1$, $S_2$ and $S_3$, these equations are the same
as one finds within the approximation of QFT on classical backgrounds. Therefore, quantum-gravity
effects (that is, the back-reaction produced by other modes) is completely encoded in the sources,
\begin{align}
 S_1 &=
\frac{G\xi^6 H^4}{k^7}\Delta{(v_{\bf k}\momv \pi_\phi^2)}
		+\frac{G\xi^4 H^2}{k^5}\sum_{{\bf q},\sigma}\left(\omega_{\mathbf{q}}^2\deltauno{v_{\bf k}\momv\vi^2}+\deltauno{v_{\bf k}\momv\momvi^2}\right),\\
		S_2&= \frac{G\xi^6 H^4}{2k^7}\left( \Delta{(\momv^2\pi_\phi^2)}- \omega_{\bf k}^2\Delta{(\vk^2\pi_\phi^2)}\right)\nonumber\\
		&+\frac{G\xi^4 H^2}{2 k^5}\sum_{{\bf q},\sigma}\left(\omega_{\mathbf{q}}^2\deltauno{\momv^2\vi^2}-\omega_{\mathbf{k}}^2\deltauno{\momvi^2v^2_{\bf k}}
		+\deltauno{\momv^2\momvi^2}-\omega_{\mathbf{k}}^2\omega_{\mathbf{q}}^2\deltauno{\vk^2\vi^2}\right),
		\\
		 S_3 &=-\omega_{\bf k}^2 S_1.
		\end{align}
Furthermore,
by combining equations \eqref{evolution_equations_4_vk},\eqref{evolution_equations_4_vkmomv} and \eqref{evolution_equations_4_momv}, it is easy to get a unique third-order equation for the fluctuation of the perturbative variable $\deltados{\vk}$,
 \begin{equation}
 	\label{equation_for_deltav}
\xi^3\frac{d^3 \deltados{\vk}}{d\xi^3}+4\xi\left(\xi^2-2\right) \frac{d \deltados{\vk}}{d\xi}+8\, \deltados{\vk}=S,
\end{equation}
where the source $S$ is defined as a linear combination of the time-derivatives of the above sources,
\begin{equation}
	S:=\xi^3\frac{d^2 {S}_1}{d\xi^2}-\frac{2\xi^3}{k}\frac{d{S}_2}{d\xi}.
\end{equation}

Our goal then is to solve equation \eqref{equation_for_deltav}. But this equation is coupled,
through the source term $S$, to other second- and fourth-order moments. Hence, in principle one should
deal with the complete system of evolution equations for the moments.
Nonetheless, since we are interested in slight modifications to the evolution
of the modes predicted by the approximation of QFT on classical backgrounds,
it is natural to consider that
the coupling between different degrees of freedom is weak,
so that they can be treated as independent variables, and thus the moments are factorized;
that is,
$\Delta(v_{\bf k}^n\pi_{\bf k}^m\pi_\phi^2)=\Delta(v_{\bf k}^n\pi_{\bf k}^m)\Delta(\pi_\phi^2)$ and $\Delta(v_{\bf k}^n\pi_{\bf k}^mv_{\bf q}^l\pi_{\bf q}^r)=\Delta(v_{\bf k}^n\pi_{\bf k}^m)\Delta(v_{\bf q}^l\pi_{\bf q}^r)$.
In addition, the moments that define the source term
$S$ will be assumed to follow the evolution given by the approximation of
QFT on classical backgrounds. These assumptions will allow us to obtain
an analytic solution to equation \eqref{equation_for_deltav}, which will provide us with insight
about the specific corrections that quantum-gravity effects produce on the power spectrum.

In particular, the factorization of the moments leads to the following form for
the sources,
\begin{align}
  S_1 &=\frac{G\xi^4 H^2}{k^5}\left[\Delta(v_{\bf k}\pi_{\bf k}) \left(\frac{\Delta(\pi_\phi^2)}{a^2}+2E_{\bf k}\right)
 + (\Delta(v_{\bf k}\pi_{\bf k}^3)+\omega_{\bf k}^2 \Delta(v_{\bf k}^3\pi_{\bf k})) \right] ,\\
  S_2 &= \frac{G\xi^4 H^2}{2k^5}\left[(\Delta(\pi_{\bf k})^2-\omega_{\bf k}^2\Delta(v_{\bf k})^2)  \left(\frac{\Delta(\pi_\phi^2)}{a^2}+2E_{\bf k}\right)
 + (\Delta(\pi_{\bf k}^4)-\omega_{\bf k}^4 \Delta(v_{\bf k}^4)) \right].
 \end{align}
In these expressions, the contribution from modes with a wavevector (and/or a sector) different from ${\bf k}$
 has been completely encoded in the function,
\begin{equation}
 E_{\bf k}:=
 -\frac{1}{2}\left[\Delta(\pi_{\bf k}^2)+\omega_{\bf k}^2\Delta(v_{\bf k}^2)\right]+\frac{1}{2}
 \sum_{{\bf q},\sigma}\left[\Delta(\pi_{\bf q}^2)+\omega_{\bf q}^2\Delta(v_{\bf q}^2)\right],
\end{equation}
which measures the difference between the energy of all modes and the energy of
the particular mode under consideration. Note that the sum is for all wavevectors
and all sectors $\sigma=s,+,\times$. Considering now the QFT evolution for
the different moments, as given in App. \ref{section_quantum_moments_QFT_classical_backgrounds}, this energy function takes the explicit form,
\begin{equation}
E_{\bf k}=-\frac{\hbar}{2} k\beta_{\bf k} \left(1-\frac{1}{\xi ^2}-\frac{1}{2 \xi ^4} \right)+
\sum_{{\bf q},\sigma}\frac{\hbar}{2} q \beta_{\bf q}\left(
 1 -\frac{k^2}{\xi^2 q^2}-\frac{ k^4}{2 \xi ^4 q^4}\right),
\end{equation}
where the factor $\beta_{\bf k}:=(2 N_{\bf k}+1)$ parametrizes the possibly excited
state of the mode $\bf k$ at the onset of inflation $(\xi\rightarrow\infty)$.
The source $S$ takes then the simple form,
\begin{equation}\label{qftsource}
 S=-\frac{4 \hbar \widetilde{H}^2 \beta_{\bf k}}{k^4 \xi^2}(\beta_{\bf k}(3+2\xi^2)+ \alpha_1 k \xi^2+9\alpha_2 k^3),
\end{equation}
where the dimensionless Hubble factor $\widetilde H:=H/\sqrt{G\hbar}$
and the positive constants,
\begin{align}
 \alpha_1 & := \sum_{{\bf q},\sigma}\frac{\beta_{\bf q}}{q}, \\
 \alpha_2 & := \sum_{{\bf q},\sigma}\frac{\beta_{\bf q}}{6\,q^3},
\end{align}
have been defined.
As a side remark, we note that the sums in the definitions of the coefficients $\alpha_1$ and $\alpha_2$
might be divergent and, in order to obtain their value,
one would need to consider a regularization. For instance, it would make
sense to consider a cut-off so that these sums only run over modes that are
actually coupled to the mode under consideration. As we have considered a fiducial volume $L^3$
of the spatial sections, the natural infrared cut-off would be given by the inverse
of this characteristic background length scale $1/L$. Regarding the ultraviolet limit,
one could consider as cut-off the inverse of the quantum-gravity scale $1/\sqrt{\hbar G}$.
In any case, for the forthcoming discussion, $\alpha_1$ and $\alpha_2$ will be considered
to be two positive constants that parametrize the quantum back-reaction
on the dynamics of a particular mode.

Let us also comment that, as can be seen in \eqref{qftsource},
under the assumption of the QFT evolution for the moments that define the source $S$,
the fluctuation of the momentum of the inflaton field $\Delta(\pi_\phi^2)$, which is
constant through evolution, has completely
disappeared from that expression. This is due to an exact cancellation of the coefficients
that multiply this term and erases all the contributions from the moments of the background
matter sector $(\phi,\pi_\phi)$ to the evolution of the fluctuation $\Delta(v_{\bf k}^2)$.
Nevertheless, even if this cancellation was not exactly fulfilled, this term would
provide a term proportional to $\widetilde{H}^4$, which would negligible,
since $\widetilde{H}$ is very small during inflation and the source $S$ is
of order $\widetilde{H}^2$.

Now, with the expression \eqref{qftsource} for the source, it is possible to obtain the analytic
solution to equation \eqref{equation_for_deltav},
\begin{align}
\Delta(v_{\bf k}^2)&=\frac{\hbar\beta_{\bf k}}{2 k \xi^2} \left(\xi^2+1\right)
+ \frac{\hbar\beta_{\bf k} \widetilde{H}^2}{6 k^4 \xi^2} \Big[11 \beta_{\bf k}+3 k \alpha_1+15  k^3 \alpha_2
\nonumber\\
&
+2 \left(\beta_{\bf k} +3 k^3 \alpha_2\right) \Big(
2 \cos (2 \xi) \left(\left(\xi^2-1\right) \text{Ci}(2 \xi)+2 \xi
   \text{Si}(2 \xi)
   -\pi  \xi \right)
   \nonumber\\
   &
-\sin (2 \xi) \left(
\left(\xi^2-1\right) (\pi -2\text{Si}(2 \xi))
+4 \xi \text{Ci}(2 \xi)\right)
   \Big)\Big],
\end{align}
where $\text{Si}$ and $\text{Ci}$ are, respectively, the sine and cosine
integral functions.
The integration constants have been fixed by requesting that at
the onset of inflation ($\xi\rightarrow\infty$) the fluctuation tends
to its stationary form $\Delta(v_{\bf k}^2)=\beta_{\bf k}\hbar/(2k)$
for a flat background\footnote{
Nonetheless, let us mention that, in the super-Hubble limit ($\xi\rightarrow 0$)
only one of the three integration constants survives and provides the amplitude
of the power spectrum ${\cal P}^{(0)}_{\bf k}$. Therefore, modifying the integration constants would
just introduce a global factor in the power spectrum ${\cal P}^{(0)}_{\bf k}$.
}. This expression shows explicitly that the trajectory of the fluctuation is
given by its form in the approximation of QFT on classical backgrounds,
as can be seen in App. \ref{section_quantum_moments_QFT_classical_backgrounds},
plus certain correction terms that are of the order of ${\widetilde H}^2$.

Considering then the definition \eqref{power_spectrum_general_adim},
and performing an expansion around the super-Hubble limit ($\xi\rightarrow 0$), 
leads to the following expression for the power spectrum
at late times,
\begin{equation}
 {\cal P}_{\bf k}={\cal P}_{\bf k}^{(0)}\left[
1
+\widetilde H^2\left(
\frac{\beta_{\bf k}}{3k^3}[11-4\gamma_E-4 \ln(2\xi)]
+\frac{\alpha_1}{k^2}
+\alpha_2[5-4\gamma_E-4\ln(2\xi)]
\right)
 \right],
\end{equation}
with ${\cal P}_{\bf k}^{(0)}:=\widetilde H^2/2\beta_{\bf k}$ being
the usual scale-independent spectrum obtained in the QFT approximation, and $\gamma_E\approx 0.577$
the Euler--Mascheroni constant.
By evaluating the numerical values, this expression can be rewritten as follows,
\begin{equation}\label{powerspectrum}
 {\cal P}_{\bf k}\approx{\cal P}_{\bf k}^{(0)}\left[
1
+\widetilde H^2\left(
\frac{1.97\beta_{\bf k}}{k^3}
+\frac{\alpha_1}{k^2}
-0.081 \,\alpha_2
-\frac{4}{3}\left(3\alpha_2+\frac{\beta_{\bf k}}{k^3}\right)\ln\xi\right)
 \right],
\end{equation}
where the time-dependence is contained in the last logarithmic term.
At the horizon crossing $(\xi=1)$ this term exactly vanishes; whereas,
from that point on, for super Hubble scales $(\xi<1)$,
it will produce a positive contribution to the power spectrum.

The expression for the power spectrum \eqref{powerspectrum} is
the main result of this paper. As can be seen,
the relative correction to the power spectrum obtained in the
QFT approximation, is proportional to $\widetilde H^2$, which
takes a very small value during inflation. In addition, this correction
depends explicitly on the wavenumber $k$ and therefore breaks
the scale-invariance of ${\cal P}^{(0)}_{\bf k}$. In fact,
concerning their dependence on the wavenumber, there are three different
contributions that encode the quantum-gravity effects.

On the one hand, the factor proportional to $\beta_{\bf k}/k^3$ comes purely from
the back-reaction of the mode with itself. That is, unlike in the
QFT approximation, the evolution of the fluctuation $\Delta(v_{\bf k}^2)$ is coupled
to higher-order moments of the same mode, like $\Delta(v_{\bf k}\pi_{\bf k}^3)$
and $\Delta(v_{\bf k}^3\pi_{\bf k})$. This coupling leads to the mentioned
contribution to the power spectrum, which is inversely proportional to $k^3$.
Therefore, this effect is more relevant for small wavenumbers, which encode
the information about large scales. Furthermore,
and as obtained also in the analysis presented in \cite{Brizuela_2019_power_spectrum_excited_states},
this correction is linear on $\beta_{\bf k}$, which parametrizes the energy of the initial state of the mode.

On the other hand, the terms that go with the positive constants $\alpha_1$ and $\alpha_2$
represent the back-reaction of all the modes on the mode under consideration.
This includes modes with a different wavenumber from the one under consideration,
as well as those from a different sector, but also self-interactions;
since the sum in the definition of the positive coefficients $\alpha_1$ and $\alpha_2$
include all of them. The term with $\alpha_2$ provides an overall shift of
the power spectrum, which is independent of the wavenumber.
Therefore, in principle this effect could be absorbed in the amplitude of the power spectrum.
Nevertheless, the term described by $\alpha_1$ is proportional
to $1/k^2$ and, like the pure self-interaction term, does include a distinctive
scale-dependence on the power spectrum.
Up to our knowledge, these two corrective factors have not been obtained on any
other previous study since, in one way or another, different computations
in the literature assume that the evolution of a given mode is not affected
by other modes. Nonetheless,
we have shown that other modes provide a correction that is of the same order of magnitude as
the one given by the mode itself, and thus their effects are not negligible.

It is interesting to note that, when particularizing the above computation to
the presence of an unique mode, the three different terms contribute
to provide a correction proportional to $\beta_{\bf k}/k^3$.
More precisely, if one considers a mode initially on its fundamental state ($\beta_{\bf k}=1$),
and neglects the contribution from other modes ($\alpha_1=1/k$,
and $\alpha_2=1/(6 k^3)$),
one obtains the power spectrum
\begin{equation}\label{ps1mode}
 {\cal P}_{\bf k}^{\rm one\, mode}={\cal P}_{\bf k}^{(0)}\left[
1
+\frac{\widetilde H^2}{k^3}\left(
\frac{11}{2}-2\gamma_E-2 \ln(2\xi)\right)
 \right]\approx{\cal P}_{\bf k}^{(0)}\left[
1
+\frac{\widetilde H^2}{k^3}\left(
2.96-2 \ln\xi\right)
 \right].
\end{equation} 
This result coincides with the one presented in \cite{brizuela_muniain_2019_moments_QG}
\footnote{In fact, there is a factor 3 of difference, which
can be explained by the fact that, in \cite{brizuela_muniain_2019_moments_QG}, a second-order
truncation in moments was considered, although quadratic combinations of second-order moments were
kept to see how higher-order contributions might enter. In this way, essentially fourth-order contributions
were estimated by quadratic combinations of second-order moments as,
for instance, $\Delta(v_{\bf k}^4)\approx \Delta(v_{\bf k}^2)^2$. But, as
can be seen in App. \ref{section_quantum_moments_QFT_classical_backgrounds}, this estimate is slightly short,
and there is indeed an exact factor 3 in that relation, that is, $\Delta(v_{\bf k}^4)=3 \Delta(v_{\bf k}^2)^2$.}
and, up to small numerical differences,
it also agrees with the form obtained in other approaches;
see in particular the most recent value found within the Born--Oppenheimer type
of approximation \cite{CK21}.
This particularization to one mode remarks the relevance of considering the contribution from
the other modes. Note that just looking at the result from one mode \eqref{ps1mode},
one would infer a correction of the form $1/k^3$ to the power spectrum.
Nevertheless, when considering other modes, some of this dependence changes
because one sums over all wavenumbers. In particular, as we have shown,
this dependence is splitted in three different terms: one that keeps
the form $1/k^3$, another one with a dependence of the form $1/k^2$ and
the third one completely independent of $k$.

\section{Conclusions}\label{section_conclusions}

In this work we have obtained quantum-gravity corrections to the power spectrum for inflationary scalar and tensor
perturbations in the context of a geometrodynamical quantization. Contrary to the usual approximation of QFT on
classical backgrounds, the quantum behavior of all the degrees of freedom of the model (both background and perturbative)
is considered in this case.
More precisely, in this paper we have generalized previous results presented in \cite{brizuela_muniain_2019_moments_QG}
in three different and relevant aspects. On the one hand, the third- and fourth-order truncations in moments have been studied.
On the other hand, initially excited states have also been included. Furthermore, and more importantly, the presence
of infinite modes have been considered. This last generalization has introduced some new scale-dependence
in the power spectrum, and has shown that the back-reaction of other modes is as important as
the self-interaction of a specific mode with itself. Therefore, as one would expect from a quantum-gravity
theory, all the degrees of freedom are coupled and it is not correct to assume an independent evolution
of a given mode, even to compute the leading-order quantum-gravity corrections.

From a methodological perspective, we have chosen the moments \eqref{q_moments_modes} as our basic variables
in order to describe the system. In this way, the
Wheeler--DeWitt equation \eqref{WdWeq} has been rewritten as an infinite set of constraint equations for these moments.
These equations form a first-class system and encode certain gauge freedom, that is related to the choice of time
variable. In our case, we have chosen the scale factor
as the internal time parameter and, thus, once the system is deparametrized, its conjugate momentum turns out
to be the physical Hamiltonian. In order to deal with this infinite system, as already commented, we have considered
two different truncations in the order of the moments.

At third order we have been able to obtain the complete expression for the physical Hamiltonian \eqref{hamiltonian_third_order_ai} for any potential.
Then, we have considered the case of a constant potential and, by constructing an initial adiabatic
state, we have concluded that third-order moments do not contribute to the power spectrum of the perturbations.

At fourth-order the analysis is technically much more involve and therefore we have assumed the constant-potential case from
the beginning. We have then obtained the evolution equation for the fluctuation of the perturbative variable
$\Delta(v_{\bf k}^2)$ \eqref{equation_for_deltav}, that is directly related to the power spectrum. The main differential part of
this equation takes exactly the same form as within the approximation of QFT on classical backgrounds,
and the quantum-gravity effects are completely encoded in the source term $S$. In order to solve this equation analytically,
we have assumed that the source term follows the evolution given by the approximation of
QFT on classical backgrounds. In this way, the exact form for the power spectrum \eqref{powerspectrum} has been computed.

As has been obtained in several other approaches in the literature, the relative correction to the power spectrum
due to quantum-gravity effects is of the order of the square of the Hubble factor $H^2$, which is a very small
quantity during inflation. In addition, quantum-gravity effects introduce a distinctive scale-dependence on the power spectrum.
In particular, there are three different terms. On the one hand, due to the self-interaction of the mode
with itself, a dependence of the form $1/k^3$ arises. On the other hand, the back-reaction of all modes
on the mode under consideration introduces two different terms: one that behaves as $1/k^2$, and another
one that is scale-independent and produces an overall shift of the amplitude of the power spectrum.
In previous studies only the dependence $1/k^3$ has been computed since, in one way or another,
the back-reaction from other modes has been completely neglected. Nevertheless, we have shown that this produces an effect
of the same order of magnitude as the self-interaction, and thus it is important to take
it into account.

\section*{Acknowledgments}
This work has been supported by Project FIS2017-85076-P (MINECO/AEI/FEDER, UE)
and Basque Government Grant No.~IT956-16.

\appendix

\section{Constraints of the form $\langle (\widehat X-X)\widehat{\cal H} \rangle$ }\label{appendix_constraints}

In this appendix we explicitly display the six independent
constraints of the form $\langle (\widehat X-X)\widehat{\cal H} \rangle$ up to third order:
\begin{align*}
	\average{(\aop-a)\mathcal{\widehat{H}}}&=-iG\hbar \moma +\frac{\momphi}{a^2}\deltauno{a\momphi}+\frac{\deltaunodos{a}{\momphi}}{a^2}-2G\moma \deltauno{a\moma}-G\deltaunodos{a}{\moma}-\frac{2\momphi^2}{a^3}\deltados{a}+\frac{4\momphi}{a^3}\deltadosuno{a}{\momphi}\nonumber\\
	&+\frac{2\momphi^2}{a^4}\deltatres{a}+8a^3\deltados{a}V(\phi)+12 a^2 \deltatres{a}V(\phi)+2a^4\deltauno{a\phi}V'(\phi)+8a^3\deltadosuno{a}{\phi}V'(\phi)\nonumber\\
	&+a^4\deltaunodos{a}{\phi}V''(\phi)+\sum_{\mathbf{k},\sigma}\left(2 \momvsigma \deltauno{a\momvsigma}+\deltaunodos{a}{\momvsigma}+2\omega_{\mathbf{k},\sigma}^2 \vksigma \deltauno{a\vksigma}+\omega_{\mathbf{k},\sigma}^2 \deltaunodos{a}{\vksigma}\right),\\
	\average{(\momaop-\moma)\mathcal{\widehat{H}}}&=\frac{i\hbar}{a^3}\left(\momphi^2-4a^6V(\phi)\right)+\frac{2\momphi}{a^2}\deltauno{\moma\momphi}
	+\frac{\deltaunodos{\moma}{\momphi}}{a^2}-G(2\moma\deltados{\moma}+\deltatres{\moma})+\frac{3\momphi^2}{a^4}\deltadosuno{a}{\moma}\nonumber\\
	&-\frac{2\momphi^2}{a^3}\deltauno{a\moma}+\frac{4\momphi}{a^3}\deltauno{a\moma\momphi}+(2a\deltauno{a\moma}	+3\deltadosuno{a}{\moma})4a^2V(\phi)+a^4\deltaunodos{\moma}{\phi}V''(\phi)\nonumber\\
	&+(a\deltauno{\moma\phi}+4\deltauno{a\moma\phi})2a^3V'(\phi)\nonumber
	\\
	&+\sum_{\mathbf{k},\sigma}\left(2\momvsigma \deltauno{\moma\momvsigma}+\deltaunodos{\moma}{\momvsigma}+\omega_{\mathbf{k}}^2(2\vksigma\deltauno{\moma\vksigma}+\deltaunodos{\moma}{\vksigma})\right),
 	\\
	\average{(\phiop-\phi)\mathcal{\widehat{H}}}&=\frac{i\hbar\momphi}{a^2}+\frac{2\momphi}{a^2}\deltauno{\phi\momphi}+\frac{\deltaunodos{\phi}{\momphi}}{a^2}-2G\moma\deltauno{\moma\phi}-G\deltadosuno{\moma}{\phi}-\frac{2\momphi^2}{a^3}\deltauno{a\phi}-\frac{4\momphi}{a^3}\deltauno{a\phi\momphi}\nonumber\\
	&+\frac{2\momphi^2}{a^4}\deltadosuno{a}{\phi}+2 a^4 \deltados{\phi} V'(\phi )+8 a^3\deltaunodos{a}{\phi} V'(\phi )+a^4 \deltatres{\phi}  V''(\phi )+8 a^3 \deltauno{a\phi} V(\phi )\nonumber\\
	&+12 a^2\deltadosuno{a}{\phi} V(\phi )+\sum_{\mathbf{k},\sigma}\left(2\momvsigma\deltauno{\phi\momvsigma}+\deltaunodos{\phi}{\momvsigma}+2\omega_{\mathbf{k}}^2\vksigma\deltauno{\phi\vksigma}+\omega_{\mathbf{k},\sigma}^2\deltaunodos{\phi}{\vksigma}\right),
	\\
	\average{(\momphiop-\momphi)\mathcal{\widehat{H}}}&=-i\hbar  a^4 V'(\phi )+\frac{2\momphi}{a^2}\deltados{\momphi}+\frac{\deltatres{\momphi}}{a^2}-2G\moma\deltauno{\moma\momphi}-G\deltadosuno{\moma}{\momphi}-\frac{2\momphi^2}{a^3}\deltauno{a\momphi}\nonumber\\
	&-\frac{4\momphi}{a^3}\deltaunodos{a}{\momphi}+\frac{3\momphi^2}{a^4}\deltadosuno{a}{\momphi}+(a\deltauno{\phi\momphi}+4 \deltauno{a\phi\momphi})2a^3V'(\phi ) \nonumber\\
	&+(2 a\deltauno{a\momphi}+3 \deltadosuno{a}{\momphi})4a^2 V(\phi )\nonumber
	\\
	&+a^4 \deltadosuno{\phi}{\momphi} V''(\phi )+ \sum_{\mathbf{k},\sigma} \left(2\momvsigma\deltauno{\momphi\momvsigma}+\deltaunodos{\momphi}{\momvsigma}+2\omega_{\mathbf{k}}^2\vksigma\deltauno{\momphi\vksigma}+\omega_{\mathbf{k}}^2\deltaunodos{\momphi}{\vksigma}\right),\\
	\average{(\widehat{\pi}_{\mathbf{q}}-\momvi)\mathcal{\widehat{H}}}&=-i\hbar\omega_{\mathbf{q}}\vi+\frac{2\momphi^2}{a^2}\deltauno{\momphi\momvi}+\frac{\deltadosuno{\momphi}{\momvi}}{a^2}-2G\moma\deltauno{\moma\momvi}-G\deltadosuno{\moma}{\momvi}-\frac{2\momphi^2}{a^3}\deltauno{\momphi\momvi}\nonumber\\
	&-\frac{4\momphi}{a^3}\deltauno{a\momphi\momvi}+\frac{3\momphi^2}{a^4}\deltadosuno{a}{\momvi}+(a\deltauno{\phi\momvi} V'(\phi )+2\deltauno{a\phi\momvi}) 4a^3V'(\phi )\nonumber\\
	&+a^4 \deltadosuno{\phi}{\momvi} V''(\phi )+(2a \deltauno{a\momvi}+3 \deltadosuno{a}{\momvi}) 4a^2V(\phi )\nonumber\\
	&+\sum_{\mathbf{k},\sigma}\big(2\vk\deltauno{\momvsigma\momvi}+\deltadosuno{\momv}{\momvi}+2\omega_{\mathbf{k},\sigma}^2\vksigma\deltauno{\vksigma\momvi}+\omega_{\mathbf{k}}^2\deltadosuno{\vksigma}{\momvi}\big),
		\end{align*}
	\begin{align*}
	\average{(\widehat{v}_{\mathbf{q}}-\vi)\mathcal{\widehat{H}}}&=i\hbar\pi_{\bf q}+\frac{2\momphi^2}{a^2}\deltauno{\momphi\vi}+\frac{\deltadosuno{\momphi}{\vi}}{a^2}-2G\moma\deltauno{\moma\vi}-G\deltadosuno{\moma}{\vi}-\frac{2\momphi^2}{a^3}\deltauno{\momphi\vi}\nonumber\\
	&-\frac{4\momphi}{a^3}\deltauno{a\momphi\vi}+\frac{3\momphi^2}{a^4}\deltadosuno{a}{\vi}+(a \deltauno{\phi\vi} +4 \deltauno{a\phi\vi})2a^3 V'(\phi )+a^4 \deltadosuno{\phi}{\vi} V''(\phi )\nonumber\\
	&+8 a^3 \deltauno{a\vi} V(\phi )+12 a^2 \deltadosuno{a}{\vi} V(\phi )\nonumber\\
	&+\sum_{\mathbf{k},\sigma}\big(2\momvsigma\deltauno{\vi\momvsigma}+\deltaunodos{\vksigma}{\momvi}+2\omega_{\mathbf{k}}^2\vksigma\deltauno{\vksigma\vi}+\omega_{\mathbf{k}}^2\deltadosuno{\vksigma}{\vi}\big).
\end{align*}

\section{Coefficients $A_0$ and $A_2$}\label{appendix_coefficients_A0_A2_third_order}

In this appendix we provide the coefficients $A_0$ and $A_2$ that appear in the physical Hamiltonian \eqref{hamiltonian_third_order_ai}
within the third-order truncation in moments:
\begin{align*}
	A_0&=\frac{1}{4} a^6  \left(V(\phi )^2 \left(a^6 \deltados{\phi} V''(\phi )+\deltados{\momphi}\right)-4 a^6 \deltauno{\phi^3} V'(\phi )^3-6 a^6 \deltados{\phi} V(\phi ) V'(\phi )^2\right)\nonumber\\
	&-\frac{\momphi^3}{16}(11 \momphi \deltados{\momphi}+16 \deltatres{\momphi})+\frac{\momphi}{16}\Bigl(\momphi \deltados{\phi}+4 \deltadosuno{\phi}{\momphi}\Bigr) V'(\phi )^2 a^6+\frac{15}{a^6}V(\phi) \momphi^2\deltados{\momphi}\nonumber\\
	&-\frac{18}{a^6}V'(\phi )\momphi^2 (\momphi \deltauno{\phi\momphi}+2 \deltaunodos{\phi}{\momphi}+18 V(\phi )\momphi \deltauno{\phi\momphi}) -\frac{3}{4a^2}\deltados{\phi} V''(\phi )\momphi^2\nonumber\\
	&+\sum_{\mathbf{k},\sigma}\Bigl\{\frac{a^8}{4} V(\phi )^2\Bigl(\deltados{\vksigma} \omega_{\mathbf{k}}^2+\deltados{\momvsigma}\Bigr)+\frac{1}{16a^4}\momphi^2 \Bigl(\Bigl(\deltados{\vksigma} \omega_{\mathbf{k}}^2+\deltados{\momvsigma}\Bigr) \momphi^2+\frac{18}{a^2}V'(\phi )\momvsigma^3 \deltauno{\phi\momvsigma}\nonumber\\
	&-24 \Bigl(\vksigma \deltauno{\momphi\vksigma}\omega_{\mathbf{k}}^2+\momvsigma \deltauno{\momphi\momvsigma}\Bigr) \momphi-2\omega_{\mathbf{k}}^2\vksigma (5 \vksigma \deltados{\momphi}-24\momv\deltaunodos{\momphi}{\vksigma}) -10 \momvsigma^2 \deltados{\momphi}\Bigr)\nonumber
	\\
	& -\frac{3}{a^4} V(\phi )\momphi \Bigl(\momphi \Bigl(\deltados{\vksigma} \omega_{\mathbf{k}}^2+\deltados{\momvsigma}\Bigr)-12 \Bigl(\vksigma \deltauno{\momphi\vksigma}\omega_{\mathbf{k}}^2+\momvsigma \deltauno{\momphi\momvsigma}\Bigr)\Bigr)\nonumber\\
	&- \frac{3}{a^4} V(\phi )\deltados{\momphi}\Bigl(\momvsigma^2+\omega_{\mathbf{k}}^2 \vksigma^2\Bigr) +\frac{1}{16a^6}\deltados{\momphi}\Bigl(\momvsigma^2+\omega_{\mathbf{k}}^2 \vksigma^2\Bigr)-\frac{3}{4a^2}\deltados{\phi} V''(\phi )\Bigl(\momvsigma^2+\omega_{\mathbf{k}}^2 \vksigma^2\Bigr) \nonumber
	\\
	&-\frac{3}{2a^2}\momphi\big(\deltauno{\momphi\momvsigma}\momvsigma^3+\momvsigma^2\big(12 \vksigma \deltauno{\momphi\vksigma} \omega_{\mathbf{k}}^2-\momphi \deltados{\vksigma} \omega_{\mathbf{k}}^2+5 \momphi \deltados{\momvsigma}+24 \deltaunodos{\momphi}{\momvsigma}\big)\big)\nonumber\\	
	&+\frac{18}{a^4}V'(\phi )\momphi\big(\momvsigma^2 \deltauno{\phi\momphi}+\momvsigma (\momphi \deltauno{\phi\momvsigma}+4 \deltauno{\phi\momphi\momvsigma})\big)-3 V(\phi )\deltados{\phi} V''(\phi ) a^2\Bigl(\momvsigma^2+\omega_{\mathbf{k}}^2 \vksigma^2\Bigr)\nonumber\\
	&+\frac{3}{a^2} V(\phi )\Bigl(-\vksigma^2 \deltados{\momvsigma} \omega_{\mathbf{k}}^2 +5\momvsigma^2 \deltados{\momvsigma}\Bigr)+36V(\phi ) V'(\phi ) a^2 \Bigl(\vksigma \deltauno{\phi\vksigma}\omega_{\mathbf{k}}^2+\momvsigma \deltauno{\phi\momvsigma}\Bigr)\Bigr\}\nonumber\\
	&+\sum_{\mathbf{k},\sigma}\sum_{\mathbf{q},\sigma'}\Big\{-\frac{3}{4a^4}\omega_{\mathbf{k}}^2\vksigma(\visigma \deltados{\phi}+4 \deltadosuno{\phi}{\visigma} +\momvisigma^2 \deltados{\phi}+4 \momvisigma \deltadosuno{\phi}{\momvisigma}) \nonumber\\
	&+\frac{18}{a^4}V'(\phi )\momphi  \omega_{\mathbf{k}}^2\vksigma \big(\momphi \deltauno{\phi\visigma}+\visigma \deltauno{\phi\momphi}+4 \deltauno{\phi\momphi\visigma}\big)- \frac{\momvsigma^3}{2} \Bigl(3 \visigma \deltauno{\vksigma\momvisigma} \omega_{\mathbf{q}}^2+2 \deltatres{\momvisigma}\Bigr)\nonumber\\
	&+\frac{3}{a^2} V(\phi )\Bigl(12 \momvsigma \visigma \deltauno{\visigma\momvsigma} \omega_{\mathbf{q}}^2+\deltados{\vksigma}\omega_{\mathbf{k}}^2\Bigl(5\omega_{\mathbf{q}}^2 \visigma^2-\momvisigma^2\Bigr)\Bigr)-\frac{\momvsigma^4}{16}\Bigl(11 \deltados{\momvisigma}-\omega_{\mathbf{q}}^2 \deltados{\visigma}\Bigr) \nonumber\\
	&+\frac{1}{8a^2}\momphi\omega_{\mathbf{k}}^2 \vksigma^2 \Bigl(12\omega_{\mathbf{q}}^2 (\visigma \deltauno{\momphi\visigma}+2 \deltaunodos{\momphi}{\visigma})-2\momphi \Bigl(\deltados{\momvisigma}-5 \omega_{\mathbf{q}}^2 \deltados{\visigma}\Bigr)\Bigr) \nonumber\\
	& -\frac{1}{16}\omega_{\mathbf{k}}^4\vksigma^3 \Bigl(11 \visigma \deltados{\visigma} \omega_{\mathbf{q}}^2+16 \deltados{\visigma} \omega_{\mathbf{q}}^2-\visigma\deltados{\momvisigma}\Bigr)\nonumber\\
	&+\frac{18}{a^2}V'(\phi )\Bigl(\momvsigma^2 \Bigl( \omega_{\mathbf{q}}^2\visigma \deltauno{\phi\visigma}+2 \deltaunodos{\phi}{\momvisigma}\Bigr)+\omega_{\mathbf{k}}^4\vksigma^2 (\visigma \deltauno{\phi\visigma}+2 \deltaunodos{\phi}{\visigma})\Bigr)\Big\}	\nonumber
	\end{align*}
	\begin{align*}
	&+\sum_{\mathbf{k},\sigma}\sum_{\mathbf{q},\sigma'}\sum_{\mathbf{r},\sigma''}\Big\{\frac{18}{a^2}V'(\phi )\momvsigma \omega_{\mathbf{q}}^2\visigma (\vjsigma \deltauno{\phi\momvjsigma}+4 \deltauno{\phi\visigma\momvjsigma})\nonumber\\
	&+\frac{3}{2a^2}\momphi \omega_{\mathbf{k}}^2 \vksigma\momvisigma (\momphi \deltauno{\visigma\momvjsigma}+\vjsigma \deltauno{\momphi\momvjsigma}+4 \deltauno{\momphi\visigma\momvjsigma})-\frac{5}{8}\momvsigma^2 \omega_{\mathbf{q}}^2\visigma \vjsigma\Bigl( \omega_{\mathbf{r}}^2\deltados{\vjsigma}+ \deltados{\momvjsigma}\Bigr)\nonumber\\
	&-\frac{3}{2} \momvsigma \omega_{\mathbf{q}}^2\visigma \left(\visigma\vjsigma \deltauno{\vjsigma\momvisigma}+\omega_{\bf q}^2 \deltaunodos{\vjsigma}{\momvisigma}+\momv \deltaunodos{\vjsigma}{\momvisigma}\right)\Big\},
\end{align*}
\begin{align*}
	A_2&=\frac{1}{8a^4}\Big[2 a^{12} V(\phi)V''(\phi)\deltados{\phi}+ a^{6}V^{\prime}(\phi)\left(  8\Delta\left(\phi \pi_\phi^{2}\right)+ a^{6} \left( 8V^{\prime \prime}(\phi) \Delta\left(\phi^{3}\right)+14 V^{\prime}(\phi) \Delta\left(\phi^{2}\right)\right)\right)\nonumber\\
	&+36 a^{6} \pi_\phi V^{\prime}(\phi) +V^{\prime \prime}(\phi) a^{6}\pi_{\phi}( \pi_{\phi} \Delta\left(\phi^{2}\right)+8\deltauno{\phi^2\momphi})+8 \pi_\phi \Delta\left(\pi_\phi^{3}\right)+2 a^6 V(\phi)\deltados{\momphi}+15 \pi_\phi^{2} \Delta\left(\pi_\phi^{2}\right)\nonumber\\
	&+\sum_{{\bf k},\sigma}\Bigl\{
	8 a^{4}  \pi_{\bf k} \Delta\left(\momphi^{2} \pi_{\bf k}\right)+15 a^{4}  \pi_{\bf k}^2 \Delta\left(\momv^{2}\right)+8 a^{2} \pi_\phi \Delta\left(\pi_\phi \momv^{2}\right)+2  V(\phi) a^8\left(\deltados{\momv}+\omega_{\bf k}\deltados{\momv}\right)\nonumber\\
	&+4 a^{8} V^{\prime}(\phi) \left(2 \Delta\left(\phi \momv^{2}\right)+7\momv\deltauno{\phi\momv}+2 \omega_{\bf k}^{2} \Delta\left(\phi \vk^{2}\right)+7\vk \omega_{\bf k}^{2} \Delta\left(\phi \momv\right)\right)+a^{4}  \momv^{2} \Delta\left(\pi_\phi^{2}\right)\nonumber\\
	&+a^{2} \pi_\phi^{2} \Delta \left(\momv^{2}\right)+28 a^{2}   \momv \pi_\phi \Delta\left(\pi_\phi \momv\right)+8 a^{2} V^{\prime \prime}(\phi) \vk \omega_{\bf k}^{2}\Delta\left(\phi^{2} \vk\right)+8 a^{4} \vk \omega_{\bf k}^{4} \Delta\left(\vk^{3}\right)\nonumber\\
	&+a^{8} V^{\prime \prime}(\phi) \left(8\momv  \Delta\left(\pi_\phi^{2} \momv\right)+\momv^2 \Delta\left(\pi_\phi^{2}\right)+\vk^{2} \omega_{\bf k}^{2} \Delta(\momphi^2)\right)+a^4\vk^2\omega_{\bf k}^2\left(\deltados{\vk}+\deltados{\momphi}+\deltados{\momv}\right)\nonumber\\
	&+a^{2} \momphi \left(8 \omega_{\bf k}^{2} \Delta  \left(\pi_\phi \vk^{2}\right)+28 \vk \omega_{\bf k}^{2} \Delta(\pi_\phi \momv)\right)+a^{2} \pi_\phi^{2}\omega_{\bf k}^2\Delta\left(\vk^{2}\right)+15 a^{4} \vk^{2} \omega_{\bf k}^{4} \Delta\left(\vk^{2}\right)
	\Bigr\}\nonumber\\
	&+\sum_{{\bf k},\sigma}\sum_{\mathbf{q},\sigma'}\Bigl\{8a^4\momv\deltauno{\momv\momvi^2}+a^{4} \momv^{2}  \Delta\left(\momvi^{2}\right)+28a^{4} \momv\left(\momvi\Delta(\momv \momvi)+\vi \omega_{\bf q}^{2} \Delta(\momv \vi)\right)+\vk^2\momv^2\deltados{\momvi}\nonumber\\
	&+a^4\vi^2\omega_{\bf k}^2\left(\momvi^2\deltados{\vk}+\omega_{\bf q}^2\deltauno{\vi^2\vk}\right)+8a^2\vk\omega_{\bf k}^2\left(\deltauno{\momphi^2\vi}+a^2\deltauno{\momvi^2\vk}\right)+28a^4\vi\vk\omega_{\bf k}^2\omega_{\bf q}^2\deltauno{\vk\vi}
	\Bigr\}
	\Big].
\end{align*}

\section{The evolution of different moments in the approximation of
QFT on classical backgrounds}
\label{section_quantum_moments_QFT_classical_backgrounds}

In this appendix we derive the evolution of the moments under
the approximation of QFT on classical backgrounds. In this case,
for each mode, one has a physical Hamiltonian and the quantum
dynamics can be written as a Schr\"odinger functional equation,
\begin{equation}
{\widehat {\cal H}}_{\bf k}\Psi_{\bf k}:= \frac{1}{2}(\hat\pi_{\bf k}^2+\omega_{\bf k}^2 \hat v_{\bf k}^2)\Psi_{\bf k}(\eta, v_{\bf k})
 = i\hbar \frac{\partial\Psi_{\bf k}(\eta,v_{\bf k})}{\partial\eta},
\end{equation}
with the conformal time $\eta$. Equivalently, one can define an
effective Hamiltonian as the expectation value of the Hamiltonian operator,
which, by performing an expansion around the expectation values $v_{\bf k}:=\langle \hat v_{\bf k}\rangle$
and $\pi_{\bf k}:=\langle\hat\pi_{\bf k}\rangle$, takes the form
$$\langle \widehat{\cal H}\rangle:=\frac{1}{2}\left(\momv^2+\omega^2_{\bf k}\vk^2+\Delta({\momv^2})+\Delta(\vk^2)\right).$$
This Hamiltonian encodes the dynamics of the different modes, and
one can obtain the equations of motion for each variable just by
computing the Poisson brackets with it. In particular, since it is quadratic,
moments of different orders completely decouple.
For the expectation values, second- and fourth-order moments one obtains the evolution equations,
\begin{align*}
	\vk'&=\momv, &\Delta(\vk^4)'&=4 \Delta(\vk^3\momv),\nonumber\\
	\momv'&=-\omega_{\bf k}^2\vk, 	&\Delta(\vk^3\momv)'&=3 \Delta(\vk^2\momv^2)-\omega_{\bf k} ^2 \Delta(\vk^4),\nonumber\\
	\Delta(\vk^2)'&=2\Delta(\vk\momv), 	&\Delta(\vk^2\momv^2)'&=2 \Delta(\vk\momv^3)-2 \omega_{\bf k} ^2 \Delta(\vk^3\momv),\nonumber\\
	\Delta(\vk\momv)'&=\Delta(\momv^2)-\omega_{\bf k}^2\Delta(\vk^2), 	&\Delta(\vk\momv^3)'&=\Delta(\momv^4)-3 \omega_{\bf k} ^2 \Delta(\vk^2\momv^2),\nonumber\\
	\Delta(\momv^2)'&=-2\omega_{\bf k}^2\Delta(\vk\momv), &\Delta(\momv^4)'&=-4 \omega_{\bf k}^2 \Delta(\vk\momv^3),
\end{align*}
where the prime stands for the derivative with respect to the conformal time $\eta$.

Taking into account the form of the frequency for the constant-potential case,
$\omega_{\bf k}^2=k^2-\frac{2}{\eta^2}$, it is immediate to solve these equations. In addition, in
order to impose the integration constants, one considers an adiabatic
state at the onset of inflation ($k\eta\rightarrow -\infty$). In particular, this
leads to a vanishing value of the expectation values,
$\vk=0$ and $\momv=0$, all along evolution, and to the following form for
the moments,
\begin{align*}
	\Delta({\vk^2})&=\frac{\hbar\beta_{\bf k}}{2k\xi^2}(\xi^2+1), &	\Delta(\vk^4)&=\frac{3 \hbar^2\beta_{\bf k}^2}{4 k^2\xi^4} \left(\xi^2+1\right)^2,\nonumber\\
	\Delta({\momv^2})&=\frac{\hbar k \beta_{\bf k}}{2\xi^4}(\xi^4-\xi^2+1), &	\Delta(\momv^4)&=\frac{3\hbar^2 k^2 \beta_{\bf k}^2}{4\xi^8}(\xi^4-\xi^2+1)^2,\nonumber\\
	\deltauno{\vk\momv}&=\frac{\hbar\beta_{\bf k}}{2\xi^3}, &\Delta(\vk^2\momv^2)&=\frac{\hbar^2\beta_{\bf k}^2}{4\xi^6}  \left(\xi^6+3\right),\nonumber\\
		\Delta(\vk^3\momv)&=\frac{3\hbar^2\beta_{\bf k}^2}{4 k\xi^5}\left(\xi^2+1\right),&\Delta(\vk\momv^3)&=\frac{3\hbar^2k\beta_{\bf k}^2}{4\xi^7}\left(\xi^4-\xi^2+1\right).
\end{align*}
These expressions have been given for the dimensionless time variable $\xi=-k\eta$.
Furthermore, the factor $\beta_{\bf k}:=(2 N_{\bf k}+1)$ parametrizes the initial
state of each mode. Finally, we are not interested in third-order moments,
as we do not need them for the computations performed in the main body of the article,
but it is easy to show that they are vanishing for the initial state under consideration.


\begin{thebibliography}{10}

\bibitem{kamenshchik_tronconi_venturi_20131_inflation_QG_BO}
A.~Y. Kamenshchik, A.~Tronconi, and G.~Venturi, ``Inflation and quantum gravity
  in a Born--Oppenheimer context'', Phys. Lett. B {\bf 726} (2013) 518.
  
\bibitem{Kamenshchik_tronconi_2014_signatures_QG_BO}
A.~Y. Kamenshchik, A.~Tronconi, and G.~Venturi, ``Signatures of quantum gravity
  in a Born{\textendash}Oppenheimer context'', Phys. Lett. B
  \textbf{734} (2014) 72.

  \bibitem{Kamenshchik:2015gua}
A.~Y.~Kamenshchik, A.~Tronconi, and G.~Venturi,
``Quantum gravity and the large scale anomaly'',
JCAP \textbf{04} (2015) 046.

\bibitem{Kamenshchik:2016mwj}
A.~Y.~Kamenshchik, A.~Tronconi, and G.~Venturi,
``Quantum cosmology and the evolution of inflationary spectra'',
Phys. Rev. D \textbf{94} (2016)  123524.

  \bibitem{Kamenshchik:2017kfs}
A.~Y.~Kamenshchik, A.~Tronconi, and G.~Venturi,
``The Born\textendash{}Oppenheimer method, quantum gravity and matter'',
Class. Quant. Grav. \textbf{35} (2018) 015012.

\bibitem{kiefer_kramer_2011_QG_contributions_CMB}
C.~Kiefer and M.~Kr\"{a}mer, ``Quantum gravitational contributions to the
  cosmic microwave background anisotropy spectrum'', Phys. Rev.
  Lett. \textbf{108} (2012) 021301.
  
  \bibitem{bini_esposito_kramer_pessina_2013_CMB_canonical_QG}
D.~Bini, G.~Esposito, C.~Kiefer, M.~Kr\"{a}mer, and F.~Pessina, ``On the
  modification of the cosmic microwave background anisotropy spectrum from
  canonical quantum gravity'', Phys. Rev. D \textbf{87} (2013) 104008.

  \bibitem{brizuela_kiefer_kramer_2016_QG_deSitter}
D.~Brizuela, C.~Kiefer, and M.~Kr\"amer, ``Quantum-gravitational effects on
  gauge-invariant scalar and tensor perturbations during inflation: The de
  sitter case'',  Phys. Rev. D \textbf{93} (2016) 104035.
  
\bibitem{brizuela_kiefer_kramer_2018_QG_slow_roll}
D.~Brizuela, C.~Kiefer, and M.~Kr\"amer, ``Quantum-gravitational effects on
  gauge-invariant scalar and tensor perturbations during inflation: The
  slow-roll approximation'',  Phys. Rev. D \textbf{94} (2016) 123527.

  \bibitem{steinwachs_vanderWild_2018_QG_wheeler_dewitt_scalar_tensor_theories}
C.~F. Steinwachs and M.~L. van~der Wild, ``Quantum gravitational corrections
  from the Wheeler{\textendash}{DeWitt} equation for scalar{\textendash}tensor
  theories'', Class. Quant. Grav. \textbf{35} (2018) 135010.

\bibitem{steinwachs_vanderWild_2019_QG_inflationary_power_spectra_scalar_tensor_theories}
C.~F. Steinwachs and M.~L. van~der Wild, ``Quantum gravitational corrections to
  the inflationary power spectra in scalar{\textendash}tensor theories'',
  Class. Quant. Grav. \textbf{36} (2019) 245015.
  
\bibitem{CK21}
L.~Chataignier and M.~Kr\"amer,
``Unitarity of quantum-gravitational corrections to primordial fluctuations in the Born--Oppenheimer approach'',
Phys. Rev. D \textbf{103} (2021) 066005.  

\bibitem{bojowald_2009_effective_constraints_rel_Qsystems}
M.~Bojowald and A.~Tsobanjan, ``Effective constraints for relativistic quantum
  systems'',  Phys. Rev. D \textbf{80} (2009) 125008.

\bibitem{bojowald_2008_effective_constraints_Qsystems}
M.~Bojowald, B.~Sandh\"ofer, A.~Skirzewski, and A.~Tsobanjan, ``Effective
  constraints for quantum systems'', Rev. Math. Phys.
  \textbf{21} (2009) 111.
  
\bibitem{BT10}
M. Bojowald and A. Tsobanjan, ``Effective constraints and physical coherent states
in quantum cosmology: A numerical comparison'', Class. Quant. Grav. \textbf{27} (2010) 145004.

\bibitem{BHT10}
M. Bojowald, P. A. H\"ohn, and A. Tsobanjan, ``An effective approach to the problem of time'',
Class. Quant. Grav. \textbf{28} (2011) 035006. 

\bibitem{BHT11}
M. Bojowald, P. A. H\"ohn, and A. Tsobanjan, ``An effective approach to the problem of time:
general features and examples'', Phys. Rev. D \textbf{83} (2011) 125023.

\bibitem{HKT12}
P. A. H\"ohn, E. Kubalova, and A. Tsobanjan, ``Effective relational dynamics of a non integrable cosmological model'',
Phys. Rev. D \textbf{86} (2012) 065014.

\bibitem{BH18}
M. Bojowald and T. Halnon, ``Time in quantum cosmology''', Phys. Rev. D \textbf{98} (2018) 066001.
  
\bibitem{brizuela_muniain_2019_moments_QG}
D.~Brizuela and U.~Muniain, ``A moment approach to compute quantum-gravity
  effects in the primordial universe'',  JCAP \textbf{04} (2019) 016.

  \bibitem{LPS97}
 J.~Lesgourgues, D.~Polarski, and A.~A.~Starobinsky,
``Quantum-to-classical transition of cosmological perturbations for non-vacuum initial states'',
Nucl. Phys. B {\bf 497} (1997) 479.
  
  \bibitem{Armendariz2007}
C. Armendariz-Picon, ``Why should primordial perturbations be in a vacuum state?'',
JCAP \textbf{02} (2007) 031.

\bibitem{Agullo2011}
I. Agullo and L. Parker, ``Non-Gaussianities and the stimulated creation of quanta in the inflationary universe'',
Phys. Rev. D {\bf 83} (2011) 063526.

\bibitem{Ganc2011}
J. Ganc, ``Calculating the local-type ${f}_{\mathrm{NL}}$ for slow-roll inflation with a nonvacuum initial state'',
Phys. Rev. D {\bf 84} (2011) 063514.

\bibitem{Aravind2013}
A. Aravind, D. Lorshbough, and S. Paban, ``Non-Gaussianity from excited initial inflationary states'',  
JCAP \textbf{07} (2013) 076.

\bibitem{Ashoorioon2014}
A. Ashoorioon, K. Dimopoulos, M. M. Sheikh-Jabbari, and G. Shiu, ``Reconciliation of high energy scale models of inflation with Planck'',
JCAP {\bf 02} (2014) 025.

\bibitem{Ashoorioon20142}
A. Ashoorioon, K. Dimopoulos, M. M. Sheikh-Jabbari, and G. Shiu,
``Non-Bunch–Davis initial state reconciles chaotic models with BICEP and Planck'',
Phys. Lett. B {\bf 737} (2014) 98.

\bibitem{Broy2016}
B. J. Broy, ``Corrections to ${n}_{s}$ and ${n}_{t}$ from high scale physics'',
Phys. Rev. D {\bf 94} (2016) 103508.

\bibitem{Brizuela_2019_power_spectrum_excited_states}
D.~Brizuela, C.~Kiefer, M.~Kr\"{a}mer and S.~Robles-P{\'{e}}rez,
  ``Quantum-gravity effects for excited states of inflationary perturbations'',
  Phys. Rev. D \textbf{99} (2019) 104007.

\end{thebibliography}
\end{document}